\documentclass[conference]{IEEEtran}
\IEEEoverridecommandlockouts
\usepackage{cite}
\usepackage{amsmath,amssymb,amsfonts}
\usepackage{algorithmic}
\usepackage{graphicx}
\usepackage{textcomp}
\usepackage{xcolor}
\usepackage{subfigure}
\usepackage{amssymb}
\def\BibTeX{{\rm B\kern-.05em{\sc i\kern-.025em b}\kern-.08em
    T\kern-.1667em\lower.7ex\hbox{E}\kern-.125emX}}

\begin{document}
%
\title{\huge DeepCP: Deep Learning Driven Cascade Prediction Based Autonomous Content Placement in Closed Social Network}
%
%
%
\author{
	Qiong Wu, Muhong Wu, Xu Chen, Zhi Zhou, Kaiwen He, and Liang Chen
\thanks{Q. Wu, M. Wu, X. Chen, Z. Zhou and K. He are with the School of Data and Computer Science, Sun Yat-sen University, Guangzhou, China, and Guangdong Key Laboratory for Big Data Analysis and Simulation of Public Opinion. L. Chen is with the Department of Financial Technology, Tencent Inc., Shenzhen, China. E-mail: \texttt{\{wuqiong23, wumh6, hekaiw\}@mail2.sysu.edu.cn}, \texttt{\{chenxu35, zhouzhi9\}@mail.sysu.edu.cn}, \texttt{leoncuhk@gmail.com}. (Corresponding author: Xu Chen)}
}
\maketitle

\begin{abstract}
Online social networks (OSNs) are emerging as the most popular mainstream platform for content cascade diffusion. In order to provide satisfactory quality of experience (QoE) for users in OSNs, much research dedicates to proactive content placement by using the propagation pattern, user's personal profiles and social relationships in open social network scenarios (e.g., Twitter and Weibo). In this paper, we take a new direction of popularity-aware content placement in a closed social network (e.g., WeChat Moment) where user's privacy is highly enhanced. We propose a novel data-driven holistic deep learning framework, namely DeepCP, for joint diffusion-aware cascade prediction and autonomous content placement without utilizing users' personal and social information. We first devise a time-window LSTM model for content popularity prediction and cascade geo-distribution estimation. Accordingly, we further propose a novel autonomous content placement mechanism CP-GAN which adopts the generative adversarial network (GAN) for agile placement decision making to reduce the content access latency and enhance users' QoE. We conduct extensive experiments using cascade diffusion traces in WeChat Moment (WM). Evaluation results corroborate that the proposed DeepCP framework can predict the content popularity with a high accuracy, generate efficient placement decision in a real-time manner, and achieve significant content access latency reduction over existing schemes.
\end{abstract}


%
\IEEEpeerreviewmaketitle

\section{Introduction}
%
%
%
%
\IEEEPARstart{W}ITH the emergence and proliferation of mobile Internet, online social networks (OSNs) such as Weibo and Twitter are undoubtedly one of the most popular Internet services. Serving as the platform for social interaction, OSNs are changing the way people discover, consume and share information and emerging as the mainstream channel for information diffusion. Inspired by this, more and more content providers (such as video content provider, HTML5 content provider, etc.) begin to offer content services on OSNs in order to expand their customer base. When a popular content (e.g., HTML5 webpage which often contains images or videos) is reshared from user to user, a content cascade can form across different regions. To maintain satisfactory quality of experience (QoE) of users for content access, a common approach is to replicate popular content close to users for low access latency. If we place the content in all regions, however, it will bring in high cache cost. If the content is only placed in a single region, users in remote regions will suffer from long access latency. In order to balance the two above, achieving efficient content placement based on content popularity in different regions is essential for the content service providers.


To address this problem, many existing studies take measurement study of content propagation pattern and social network analysis to optimize the content placement strategy. For example, Wang et al.  \cite{EnhancingWang} investigate the relationship of number of views with root users, re-share users and influenced users in content cascade, respectively. They also analyze the social relationship between users and find that video content popularity is highly affected by the online social interactions. Based on these analyses, they design a neural network-based learning framework to predict the content popularity and then propose a proactive video placement algorithm to improve video download performance. In \cite{HuWCWHZW14}, Hu et al. represent the connection between two users by their social relationship, geo-distance and preference similarity and thus the social network can be regarded as a weighted graph. Then they classify users into a collection of communities and propose a community based video content placement method. These approaches are generally suitable for open social networks such as Weibo and Twitter. Open social networks are the most common form of online social networks (OSN), where users' profiles and social connections are publicly accessible by default. This type of social networks mostly serves as a platform for snippets of interesting news and conversation flows on socially popular topics, but it may lead to information overload and privacy problems.

On the other hand, there is another kind of online social networks playing an indispensable role in people's daily life, called closed social networks (CSN), which are based on strong social ties of users and offer enhanced privacy for users, such as WeChat Moment (WM), LINE, QQ, etc. In these social networks,
a content provider (e.g., HTML5 page provider) can only observe the content diffusion trajectory formed by the set of users with reshare or view  behaviors, instead of the whole network structures including users who ignore the spreading content. In addition, the content provider is generally not allowed to access to users' personal profiles and their social relationships, which makes existing methods for cascade prediction and content placement not applicable for closed social networks \footnote{It can be also highly controversial for the OSN provider (e.g., Facebook) to immoderately utilize users' privacy-sensitive personal and social information for business purpose. In this regard, privacy-friendly approach can be highly desirable from the OSN operator's perspective.}. Furthermore, traditional model-based approaches can hardly well capture the complex content  cascade dynamics and fails to accurately predict the bursting effect particularly at the early stage, which hinders achieving efficient proactive content placement. Also traditional optimization based approach would suffer the severe scalability issue due to the huge decision space in practice.


To overcome these drawbacks and achieve intelligent service management, in this paper we propose a novel AI-enabled data-driven framework for efficient CSN content popularity prediction and autonomous content placement without utilizing users' personal and social information. Such paradigm is becoming feasible with the great support by the emerging key techniques such as deep learning, network function virtualization and software-defined networking, which are driving the network management and resource orchestration into a more smart, flexible and agile fashion \cite{wu2019mobile, zhou2019edge, li2019edge}. Specifically, for accurate cascade prediction, we propose a novel time-window Long Short-Term Memory (TW-LSTM) mechanism which captures the cascade diffusion sequences by LSTM and introduces a new time-window structure to the learning module in order to learn from the dynamic content propagation behavior data at the early stage. Accordingly, we then devise a real-time content placement algorithm CP-GAN by leveraging the power of Generative Adversarial Network (GAN). As content placement decision problem is by nature a combinatorial optimization problem which is usually time-consuming, we first leverage the sample generation ability of GAN to learn from the limited optimized decision samples which are obtained in an offline manner, and then use the discriminator of GAN to generate efficient online content placement decisions in a real-time manner. We  introduce novel design to GAN such that the discriminator of GAN has two classifiers: one is to judge whether the output is a feasible solution, and the other considers the placement as multiple binary decisions to decide which regions to place the content. As demonstrated by our experiment, CP-GAN can achieve excellent sample generation and wise content placement decision making after training with limit optimal decision samples.

We summarize the main contributions of this paper as follows:

\begin{itemize}		
	\item We advocate a novel DeepCP framework which conducts diffusion-aware cascade prediction and autonomous content placement in a holistic manner by leveraging the toolset of deep learning. The DeepCP framework is privacy-friendly and does not rely on users' personal and social information.
	
	\item To predict content popularity, we first analyze the diffusion pattern of content cascades from infection, topological and geographical perspectives. Based on the analysis, we observe the cascade diffusion process at early stages and extract features from both macro and micro views. We then design a time-window LSTM (TW-LSTM) method which can learn the evolution of cascade diffusion and well predict the eventual content popularity and geo-distribution.
	
	\item We devise a novel autonomous content placement decision generation mechanism with the predicted geo-distribution of content popularity when the content cascade is predicted to burst by TW-LSTM. The mechanism, called CP-GAN, first solves the offline content placement optimization problem for optimized decision sample acquisition and then adopts the generative adversarial network (GAN) model for efficient online content placement decision generation. By doing so, we can achieve real-time content placement decision making, without invoking time-consuming optimization solver at run time.
	
	\item Extensive experiments are conducted using a realistic closed social-network dataset, which demonstrates the high efficiency of our proposed DeepCP framework. More precisely, TW-LSTM method can achieve more than $9\%$ accuracy improvement even when comparing with neural network method for content popularity prediction. With the predicted geographical distribution of content popularity, CP-GAN can well approximate the offline optimal placement decisions, and significantly outperform existing popular approaches such as greedy scheme. As corroborated by our evaluation, CP-GAN has a swift running time and is insensitive to the input system size, which is essential for proactive content placement of fast-spreading cascades.
	
\end{itemize}

The rest of this paper is organized as follows:  In Section \ref{SectionRelated}, we review the related work of content popularity prediction and content placement decision. Section \ref{SectionAnalysis} analyzes the phenomenon of cascade diffusion from multiple perspectives and outlines the framework overview. In Section \ref{SectionCP1}, we describe the details of our TW-LSTM method which predicts the burstness of cascades and furthermore estimates the eventual geo-distribution of content popularity for explosive cascades. In Section \ref{SectionPlacement}, we introduce our proposed GAN based autonomous content placement decision generation mechanism CP-GAN. We conduct experiments in Section \ref{SectionExper}.  Our final conclusion is given in Section \ref{SectionConc}.

\section{Related Work}
\label{SectionRelated}
In this section, we describe prior works in content popularity prediction and content placement, and then highlight the key differences of our approach from existing studies.
\subsection{Content Popularity Prediction}
There are two main approaches for content popularity prediction: static information based methods and dynamic information based methods. For the first category, Wang et al. \cite{Wang2012Propagation} summarize important characteristics from social, geographical and temporal localities and use these as content popularity indices. Hu et al. \cite{HuWCWHZW14} explore the geographical and temporal diversity of content popularity and conduct a community based popularity measurement. In \cite{BastugBD14}, content popularity is predicted through collaborative filtering by exploiting user-file correlations and users' social relationships. In \cite{ZhangLZZLWW17}, Zhang et al. predict content popularity by taking the interest and behavior of users into consideration.

For the dynamic information based approaches, the content popularity prediction is considered as cascade prediction, which can be categorized into point process based methods and feature construction methods. Shen et al. \cite{Shen2014Modeling} propose a probabilistic framework to model and predict popularity dynamics based on a reinforced Poisson process. In \cite{Zhao2015SEISMIC}, the authors combine post infectiousness and human reaction time into the proposed SEISMIC model to simulate the formation of an information cascade. Different from these methods, feature construction methods usually extract rich features from multiple perspectives after having observed a cascade at the early stage \cite{Cheng2014Can, Kustarev2012Prediction, Ma2014On}. For example, Cheng et al. \cite{Cheng2014Can} represent a cascade by five classes of features (content features, original poster features, resharer features, structural features and temporal features) and then use machine learning classifiers to predict its future size.

These existing approaches all involve users' personal profiles and social graph information. For instance, Song et al. \cite{song2005modeling} predict human behavior of disseminating information by analyzing the contact and content of personal communications. In \cite{guille2012predictive}, the authors highlight that all ties and friendships in the network should be observed in their predictive model. Nevertheless, these inforamtion is usually privacy-sensitive and difficult to access to by content providers in closed social networks. In this paper, we hence propose a novel time-window LSTM model which only utilizes the observable propagation information of content cascade to estimate content popularity.

\subsection{Content Placement Decision}
Given the massive number of users in OSNs, it is important to study the popular content placement for enhancing users' QoE. In general, most studies address this problem by applying greedy or heuristic algorithms \cite{Salahuddin2015Social, Hu2013Practical, Chen2012Intra, Luss2009Optimal, PapagianniLP13}. For example, Hu et al. \cite{Hu2013Practical} propose a content placement algorithm called Differential Provisioning and Caching algorithm, which aims to give a good content-caching decision so that the total cloud-renting cost can be minimized and all demands are served.

Taking content popularity into consideration, in \cite{EnhancingWang}, the authors first take a measurement study to exploit influential factors for predicting content popularity and then design a proactive video content deployment algorithm based on prediction results. Zhu et al. \cite{Zhu2018Geo} analyze the information diffusion patterns and content is cached in locations where many users are interested in the information. In \cite{TanzilHK17}, a mixed integer linear programming problem is constructed to compute where to place content. These algorithms above are usually obtained by solving optimization problems with estimated content popularity. However, it is challenging to solve these content placement problems, which are usually NP-hard, even with an accurate prediction on the content popularity.

Different from existing works, in this paper we propose a novel autonomous decision generation mechanism CP-GAN which learns decision strategy from the offline optimal content placement samples by leveraging the power of GAN and generates efficient online content placement decision in a real-time manner at run time. To the best of our knowledge, our work is the first thrust to leverage GAN for addressing content placement problem.

\section{Data Analysis \& Framework Overview}
\label{SectionAnalysis}
\subsection{Dataset Description}
\label{SubsectionDataset}

Our study in this paper is based on a real-world dataset which contains 300 million records (35 million reshares and 265 million views) with 2250 webpages spreading in WeChat Moment (WM) in July 2016. WM, a function of China’s largest acquaintance social networking platform
— WeChat, allows users to create, view and share webpages in HTML5 (WM pages) among
friends. In this kind of closed social network, two friends mutually follow each other if they both approve the other's request, resulting in strong social ties among users. Moreover, two users in WM cannot see each others' posted contents if they are not connected as friends, indicating that only a user's friends are able to view or reshare the posted contents. The limit that each WM user can only have no more than 5000 friends further restricts on the information diffusion process. The dataset is provided by FIBODATA \footnote{http://www.fibodata.cn}, a company that provides service for creating and caching multimedia WM page contents. In this dataset, only the first reshare or view behavior of users will be recorded and all recorded users are anonymised by user indexes to preserve privacy. Moreover, as WM is a closed social network, user's personal information (such as preference, historical behavior, locations) and friendship relationships are not available, which is quite different from studies in open social networks. This also implies that the proposed algorithm in this paper is privacy-friendly.

\begin{figure}[!t]
	\centering
	\includegraphics[width=0.9\linewidth]{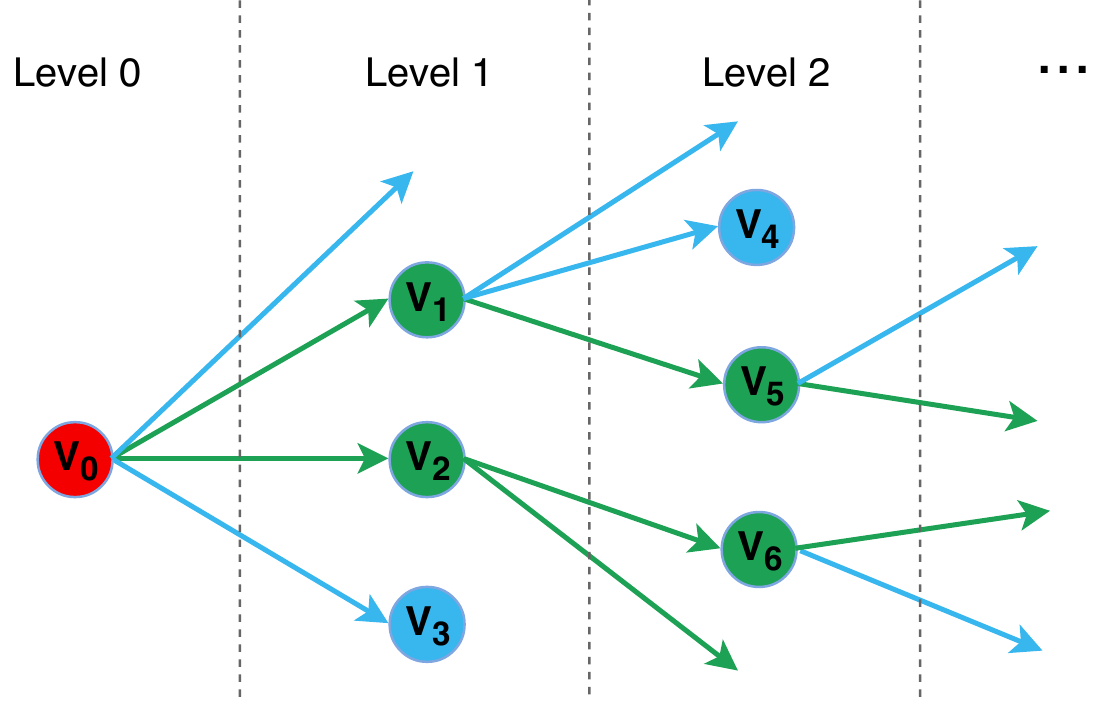}
	\caption{An example of information diffusion cascade. Blue lines indicate view behavior while green lines represent reshare behavior. The cascade grows in the form of a tree and the level of a node is its depth in the tree.}
	\label{cascade_example1}
\end{figure}

The information spread in WM is mostly first posted by WeChat Official Account (WOA), which enables businesses to continuously deliver campaign information and let followers take initiative to publicize for their company across their circles of friends through WeChat moments with low dissemination cost and high propagation rate. As described in Fig. \ref{cascade_example1}, a content (a WM page) diffusion process can be viewed in the form of cascade. The node $V_{0}$, usually known as WOA (WeChat Official Account), firstly publishes a WM page and broadcasts to its followers. Some of these followers (e.g., $V_{1}$, $V_{2}$) reshare the WM page to their own friends while others (e.g., $V_{3}$) only view the page (without resharing), which will make the information unaccessible to their friends. As the WM page is reshared one by one, the information will be spread across the social network. If we consider the cascade diffusion process as a directed tree, then the level of a node is its depth in the tree. For example, user $V_{1}$ is in level $1$ because she shares the WM page from $V_{0}$, the root of the cascade tree, whose level is $0$. The cascade propagation size is the number of nodes (users) involved in the cascade tree, which is the total number of views and reshares. The webpages in our dataset can be represented as this kind of cascades, which describes the diffusion process of contents. 

\subsection{Empirical Analysis}
As content is spread in the form of cascade, it is critical to predict whether it can burst so that we can decide whether to replicate in order to guarantee user's QoE. We first give the definition of cascade explosion and then conduct a data-driven analysis from infection, topological and geographical perspectives and the content type of pages, which identifies factors that can be beneficial for cascade burst prediction. Meanwhile, the geographical perspective can also provide some insights for where to replicate content.

\begin{figure}[!t]
	\centering
	\includegraphics[width=0.9\linewidth]{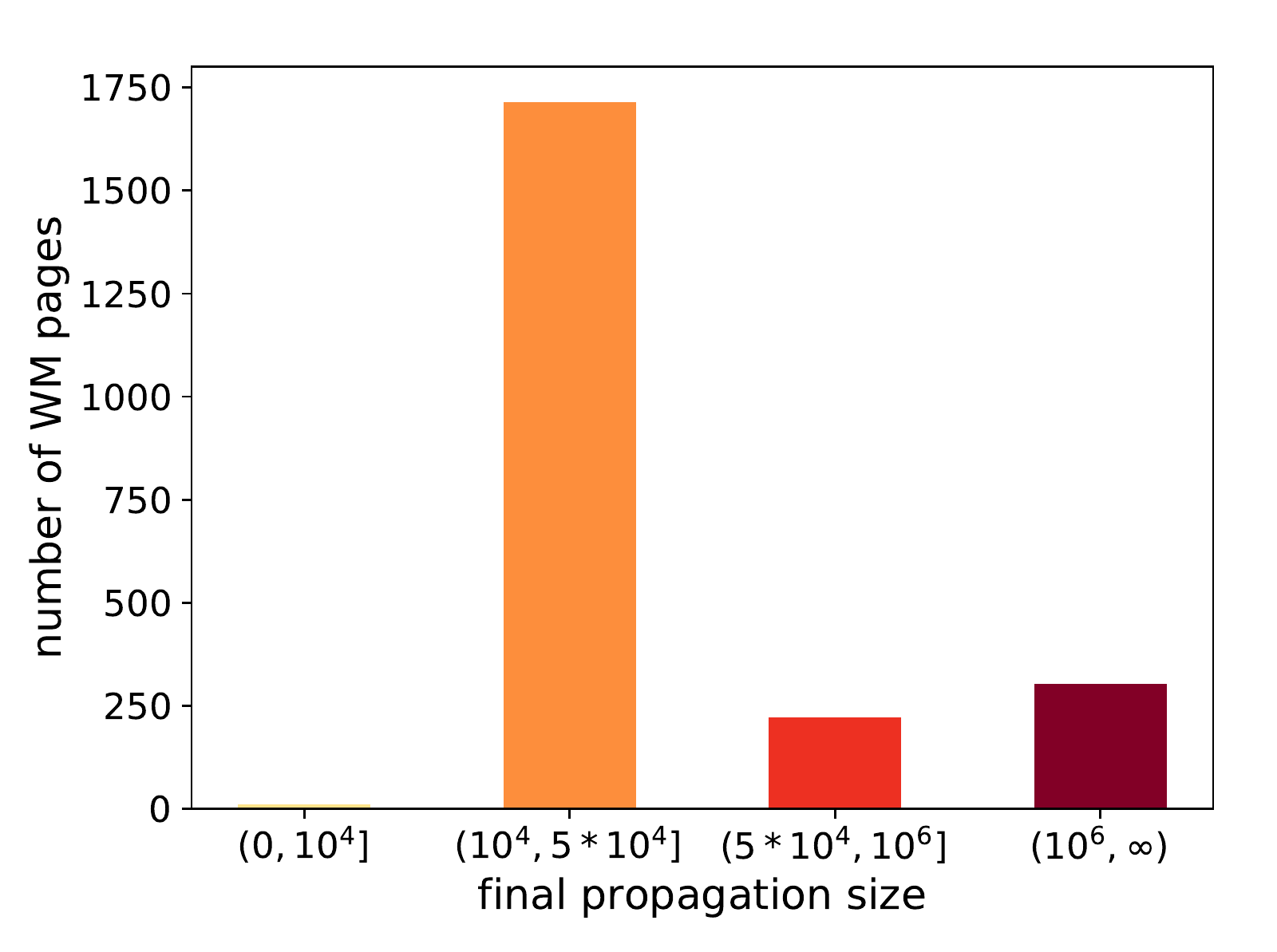}
	\caption{The cascade size distribution. Cascades differ greatly in size. Most cascades' size is between 10,000 and 50,000. Only a few cascades can infect more than 50,000 users.}
	\label{explosion_define}
\end{figure}

\begin{figure*}[!t]
	\centering
	\subfigure[cascade diffusion]{
		\includegraphics[width=0.22\linewidth]{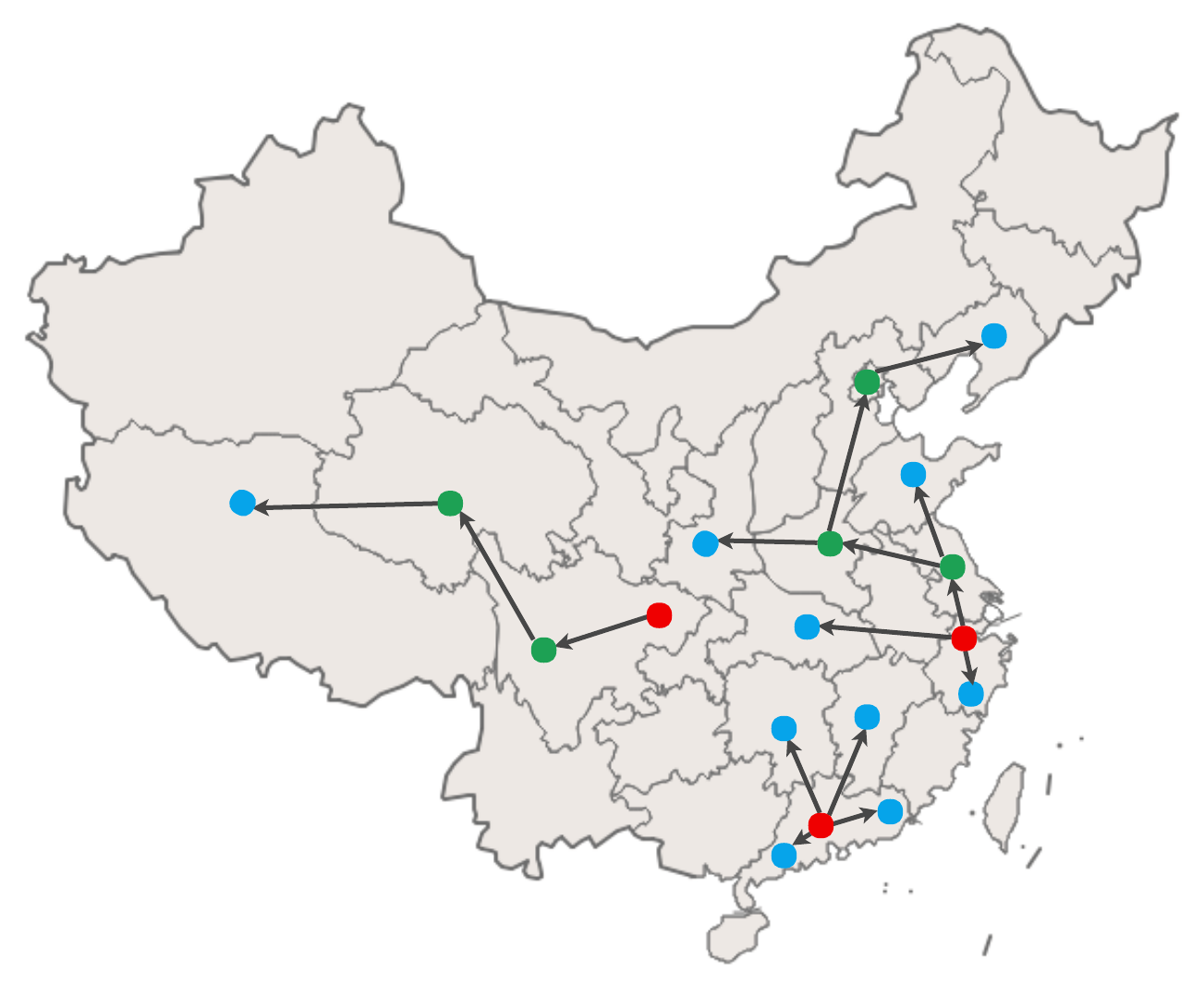}
	}
	\subfigure[infection perspective]{
		\includegraphics[width=0.22\linewidth]{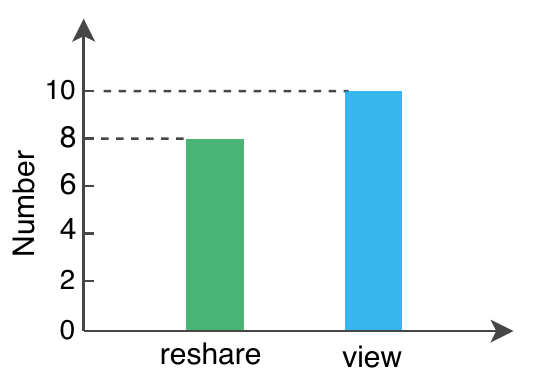}
	} 	
	\subfigure[topological perspective]{
		\includegraphics[width=0.22\linewidth]{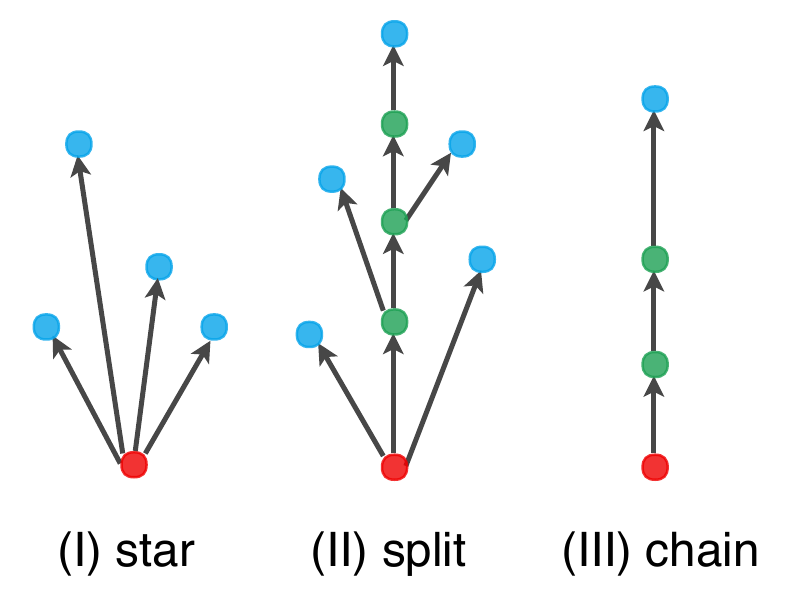}
	}
	\subfigure[geographical perspective]{
		\includegraphics[width=0.22\linewidth]{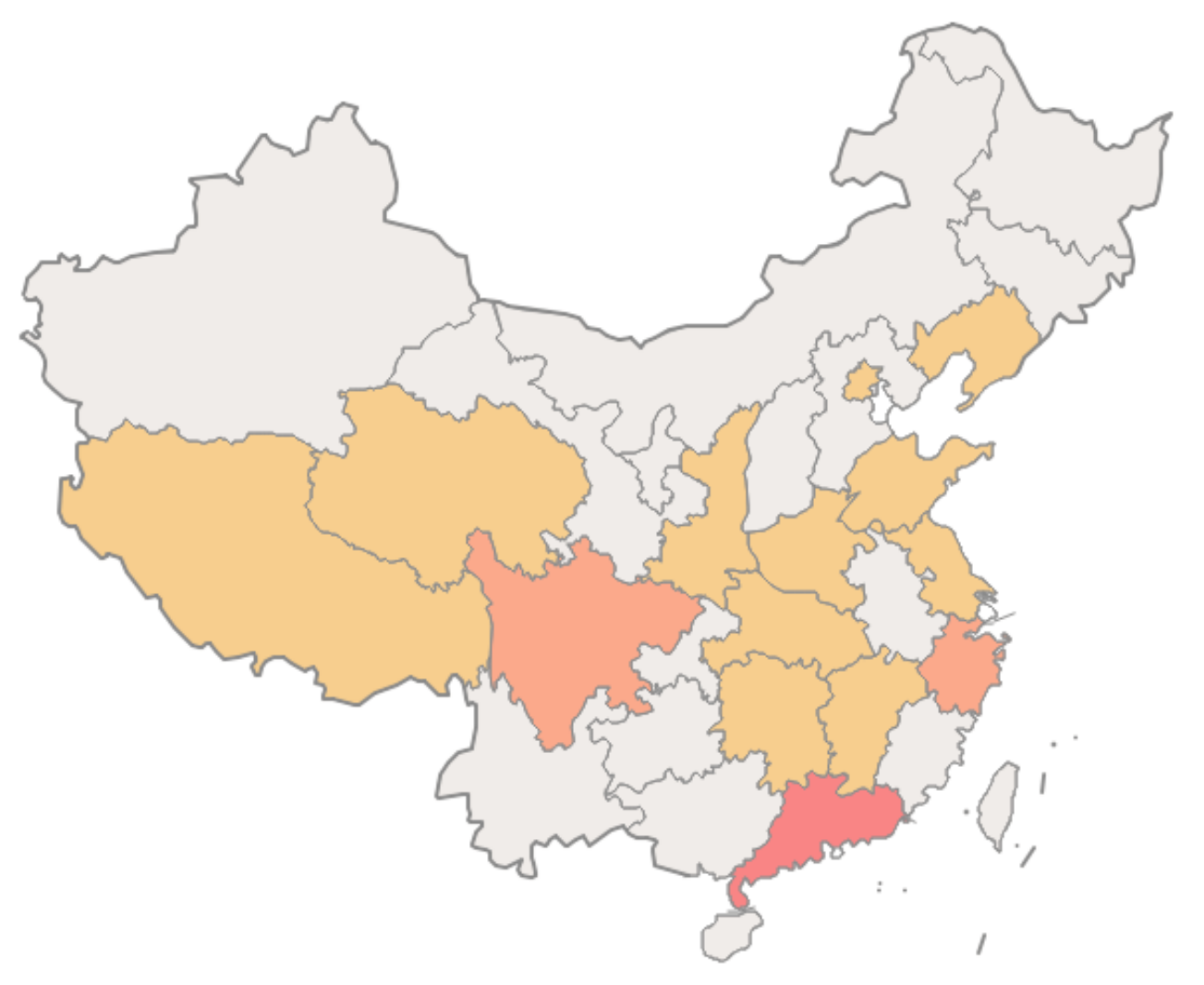}
	}
	\caption{Analysis for cascade diffusion process. (a) shows geo-spatial diffusion process in a time window. The view number and reshare number infected in this time window are listed in (b). As shown in (c), the cascades diffused in (a) can be grouped into three types: ``star'' pattern, ``split'' pattern and ``chain'' pattern. From geographical perspective, the hot areas of cascade diffusion process in (a) are highlighted in (d).}
	\label{cascadeAnalysis}
\end{figure*}

\textbf{Cascade explosion.}
As illustrated in Fig. \ref{explosion_define}, most WM pages can infect 10,000 to 50,000 users. Only $24\%$ of cascades in our dataset can reach size over 50,000 and therefore we consider these cascades to be explosive \footnote{Our methods can still be applied when other thresholds for cascade explosion are adopted.}. In our study, given the snapshot(s) of diffusion patterns at initial stages of a cascade, we aim to identify whether the cascade can grow to large scale (i.e., beyond the propagation size threshold 50,000) in the end.

\begin{figure*}[!t]
	\centering
	\subfigure[Final propagation size versus number of views infected.]{
		\includegraphics[width=0.23\linewidth]{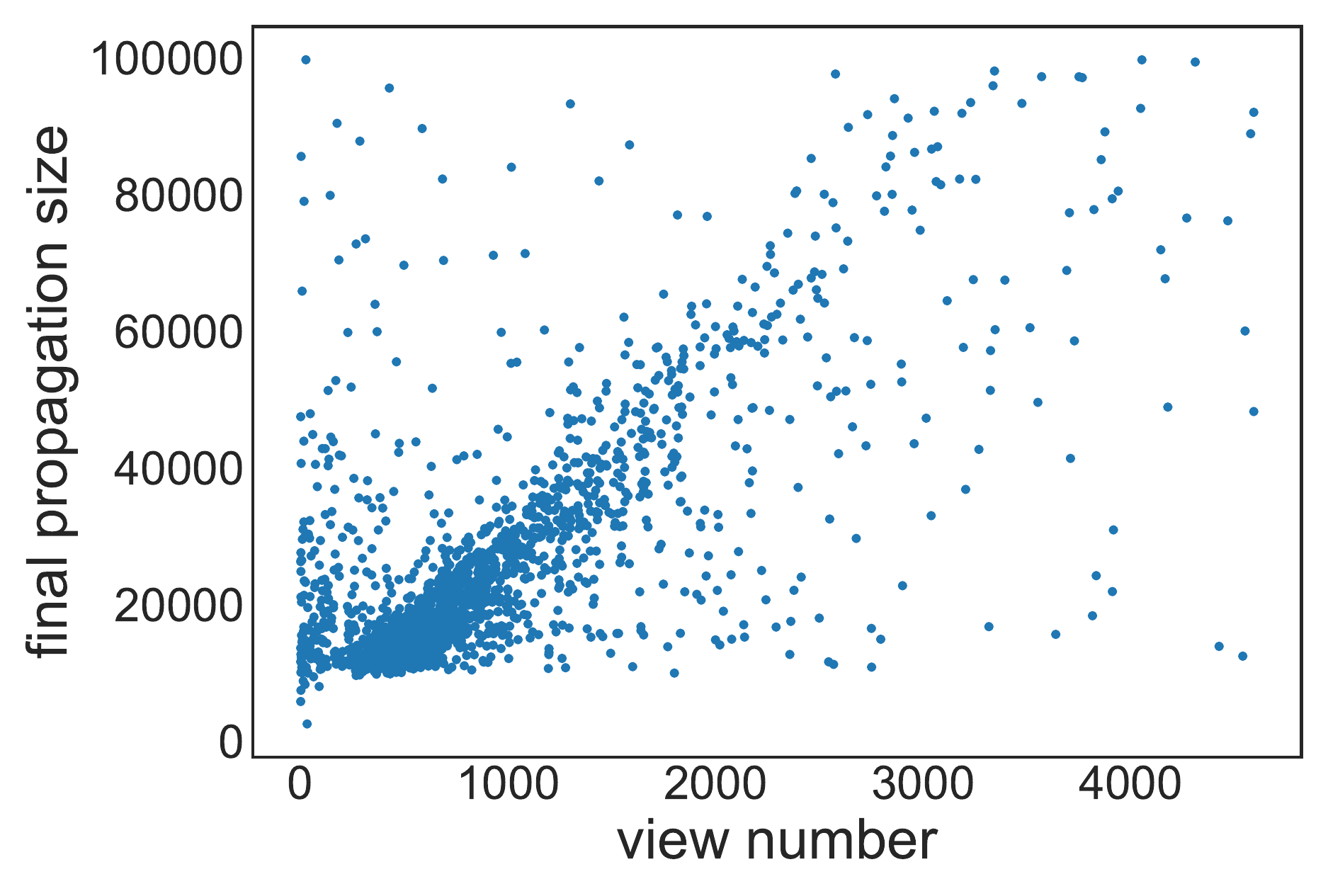}
	}
	\subfigure[Final propagation size versus number of reshares infected.]{
		\includegraphics[width=0.23\linewidth]{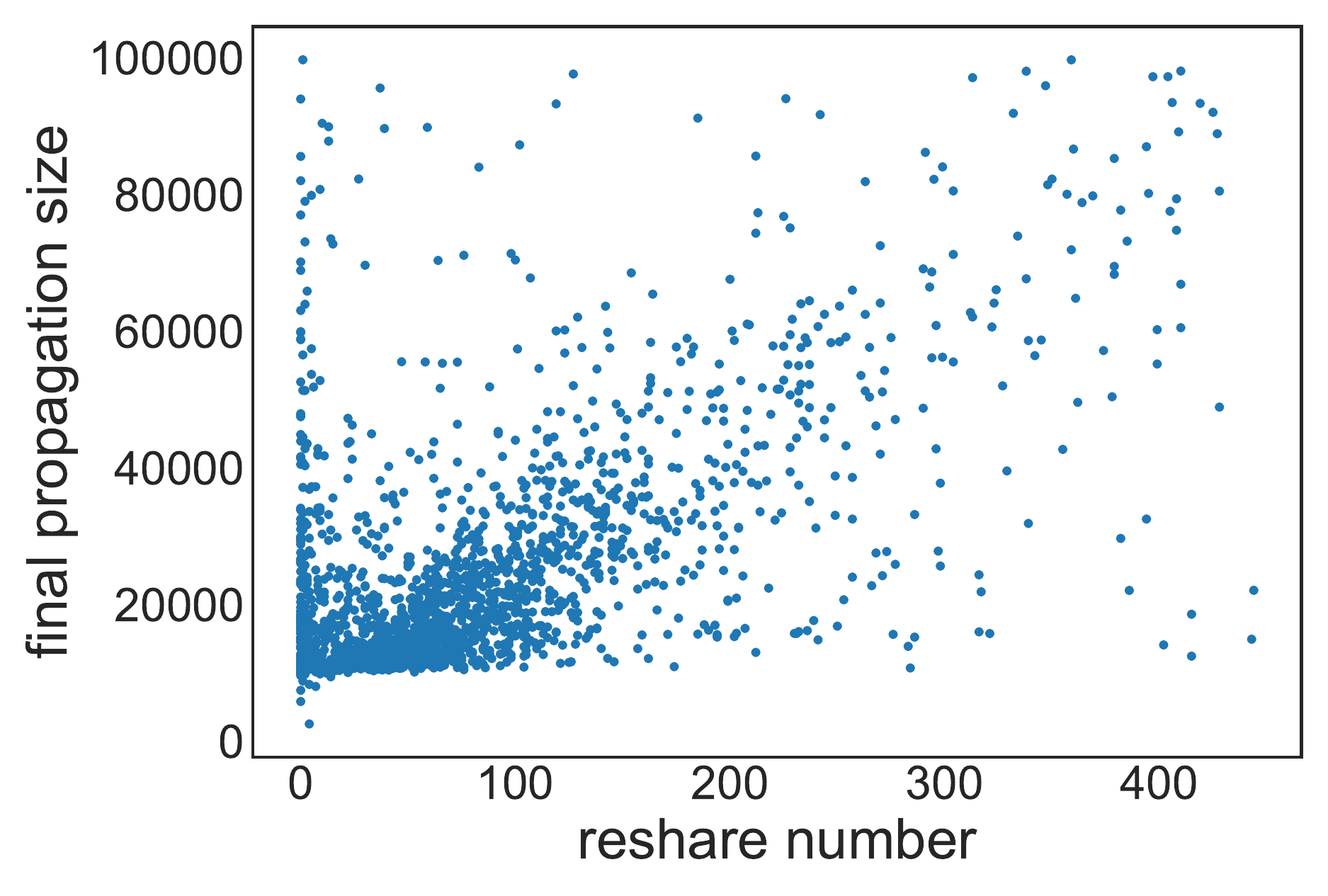}
	} 	
	\subfigure[Final propagation size versus average number of views for each level.]{
		\includegraphics[width=0.23\linewidth]{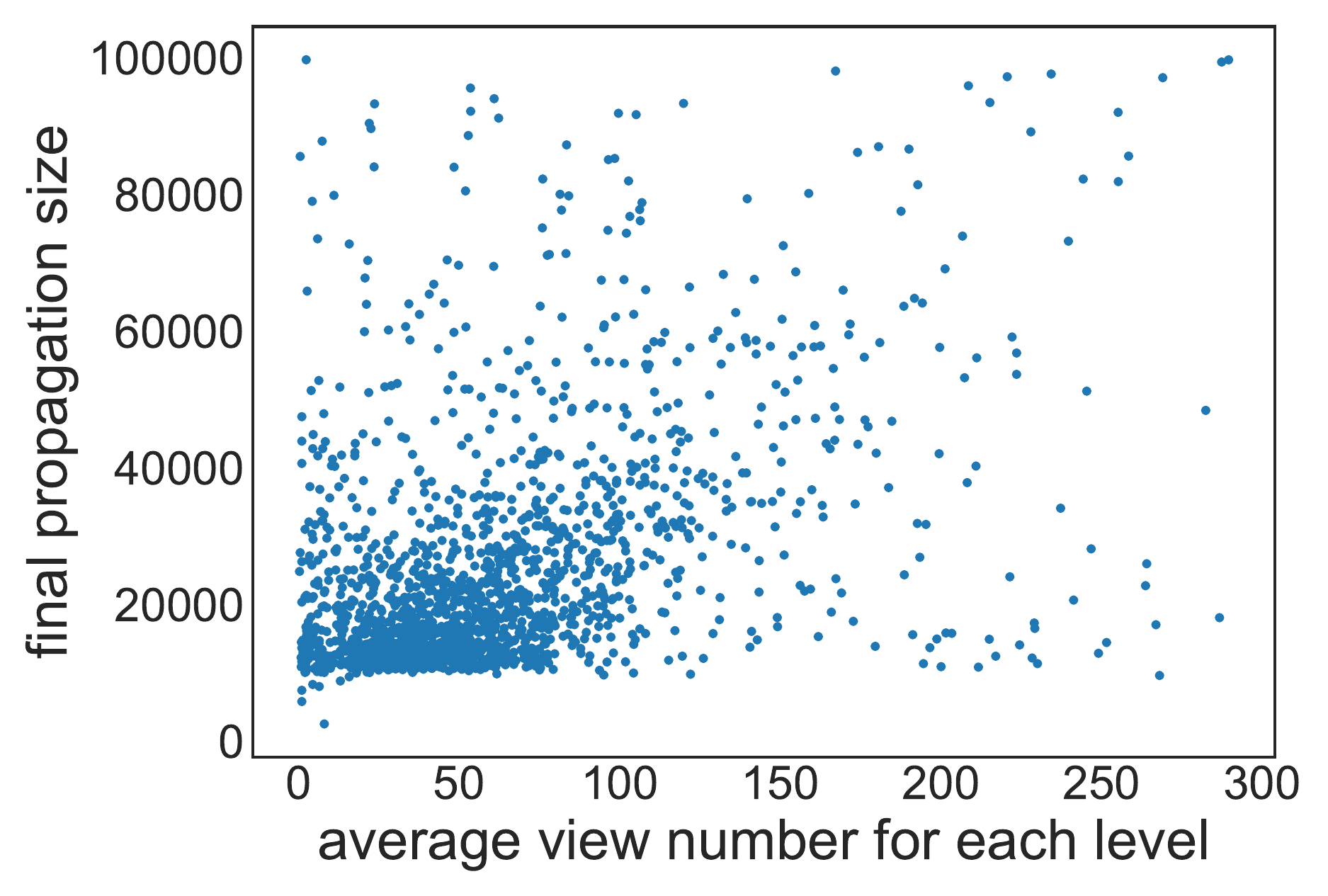}
	}
	\subfigure[Final propagation size versus taverage number of reshares for each level.]{
		\includegraphics[width=0.23\linewidth]{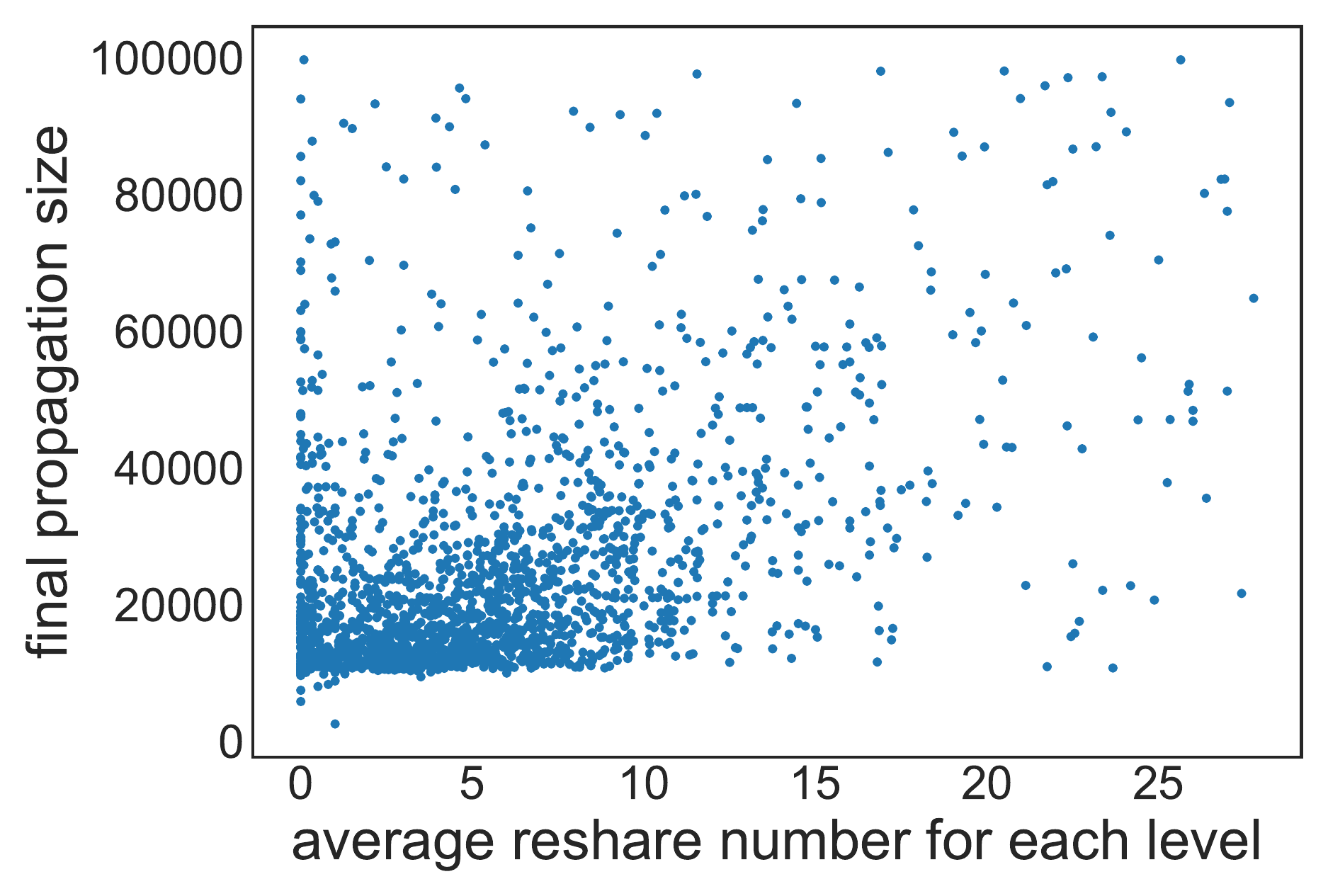}
	}
	\caption{Data-driven analysis of factors affecting the final propagation size (the total number of views and reshares) of cascades in the first time window.}
	\label{statistics}
\end{figure*}

\textbf{Infection perspective.} For a given time window (e.g., one day), the infection ability of a cascade can be measured by the numbers of views and reshares happened during this period of time. For example, Fig. \ref{cascadeAnalysis} (a) shows a WM page cascade diffusion process in a time window. Red nodes are users who have been infected (i.e., those users who have reshared the WM page) in previous time and are the source of information in this time window, green nodes represent users who reshare the WM page while blue nodes stand for users who only view it without resharing. Fig. \ref{cascadeAnalysis} (b) lists the numbers of views and reshares extracted from the diffusion process in Fig. \ref{cascadeAnalysis} (a). We can observe that the final propagation size of a cascade (the total number of views and reshares) is proportional to its infected views or reshares in the first time window to a certain degree in Fig. \ref{statistics} (a) and (b).

\textbf{Topological perspective.} The topological properties of a cascade can reflect the information dissemination pattern to some extent \cite{Liu2017Spatio}. Fig. \ref{cascadeAnalysis} (c) illustrates three representative topological structures of cascades spread in Fig. \ref{cascadeAnalysis} (a). The ``star'' pattern indicates that broadcast plays an important role in information diffusion process. This kind of cascades tend to grow wider and wider as the propagation level increases and as a result, they are more likely to be explosive. While ``chain''-pattern cascades imply that the content information is not very popular and may result in fewer people being exposed to it. As a combination of the two above, ``split'' pattern is most common in the cascades. To evaluate the topological properties of different patterns of cascades in a general way, we extract the maximum, minimum, average levels of the cascades and their variance in a specific time window as these features can reflect the structure of cascades. We can further combine level information with infection statistics (the number of views and reshares in a time window), which may be beneficial for cascade burst prediction. As illustrated in Fig. \ref{statistics} (c) and (d), the final propagation size of a cascade is related to the average number of views/reshares in each level in the first time window to a certain degree.

\textbf{Geographical perspective.} As shown in Fig. \ref{cascadeAnalysis} (a), each infected user in cascade diffusion process belongs to a specific region. As information travels across different regions when users reshare to their friends, we can obtain users' distribution among different regions in geographical space. Note that the geo-location information is coarse-grained at the city or province level and can be obtained based on users' network IP addresses by the content provider. For example, Fig. \ref{cascadeAnalysis} (d) shows the heat map of the distribution of infected users as information spread in Fig. \ref{cascadeAnalysis} (a) in a given time window. A darker color represents a larger infected user size in the region. It depicts the content popularity in different regions, which may be helpful for content placement. However, as information propagates among regions in China over time, the geographical distribution of infected users changes as well.

\begin{figure}[!t]
	\centering
	\includegraphics[width=0.9\linewidth]{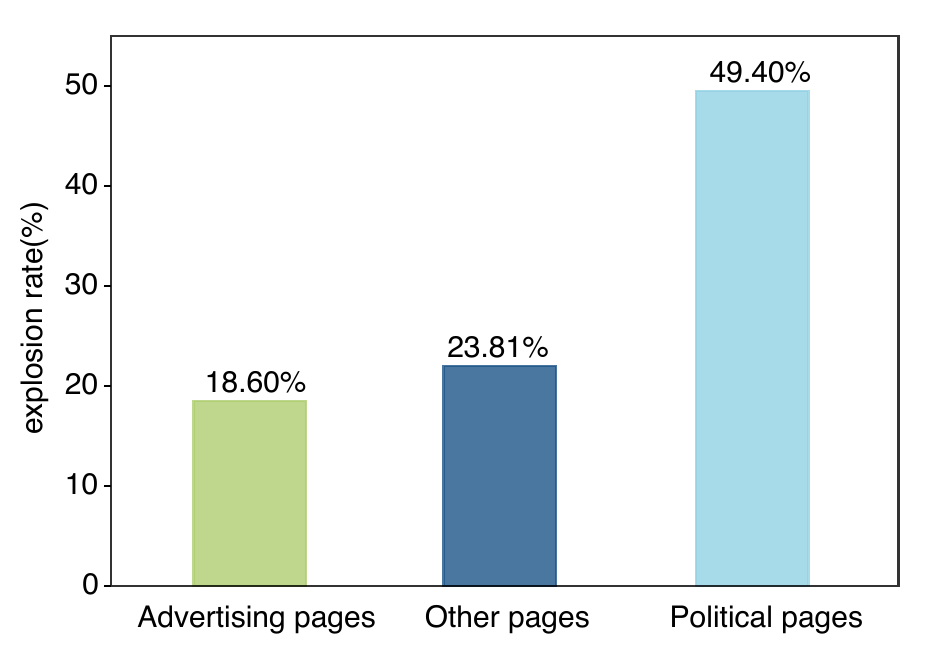}
	\caption{The explosion rates for different types of WM pages. $49.40\%$ of political pages can infect more than 50,000 users while only $18.60\%$ of advertising pages can burst.}
	\label{ad_pol}
\end{figure}

\textbf{Content Type.} Pages propagated in WM can have different content types and in turn, could result in various cascade sizes. Political and advertising pages are the two most common types of WM pages as WM provides an ideal platform for spreading public opinion and maximizing market effectiveness. Due to this fact, to study the impact of content types on cascade explosion, we compare the explosion rates of political, advertising and other kind of pages as illustrated in Fig. \ref{ad_pol}. We can find that political pages are more likely to burst while only $18.60\%$ of advertising pages can infect over 50,000 users, which may contribute to cascade burst prediction.

\subsection{Framework Overview}
\begin{figure}[!t]
	\centering
	\includegraphics[width=0.9\linewidth]{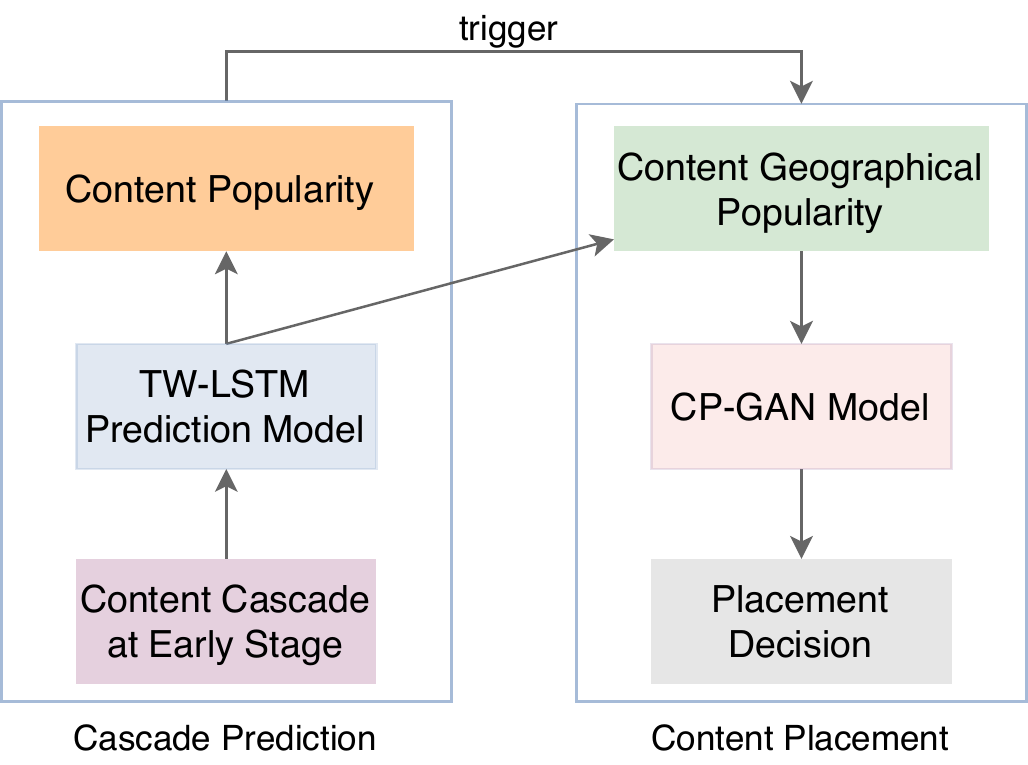}
	\caption{The framework overview.}
	\label{overview}
	\vspace{-10pt}
\end{figure}

Motivated by the data-driven analysis above, we aim to provide a holistic framework for joint popularity prediction and content placement for WeChat Moment. As described in Fig. \ref{overview}, to predict content popularity, we propose a time-window LSTM (TW-LSTM) prediction model which can learn from the evolution of content cascade at an early stage without utilizing users' personal information and social relationships. The design details of TW-LSTM model will be elaborated later in Section \ref{SectionCP1}. When a content cascade is predicted to burst, it will evaluate the content geographical popularity in the future and trigger the execution of content placement. To get the best content replication strategy in a real-time manner, we resort to the powerful generative adversarial networks (GAN) for learning efficient decision patterns and design a CP-GAN model which maps content popularity in different regions to content placement decisions. We will describe the autonomous content placement based on the predicted content popularity in Section \ref{SectionPlacement}. As we take advantage of deep learning in both cascade prediction (CP) and content placement (CP), we unify the deep learning modules into a holistic framework called DeepCP as shown in Fig. \ref{Framework}.

We should emphasize that the proposed framework is not only applicable to closed social networks, such as WM, LINE and QQ, where user's personal information and social correlations are unavailable, but can also be applied to open social networks by easily adding extra information (e.g., content freshness, user's social relationship) as the new input features in the learning modules.

\section{Diffusion-aware Cascade Prediction}
\label{SectionCP1}

In this section, we provide details for our proposed TW-LSTM method for diffusion-aware cascade prediction to estimate content popularity. This method contains two parts: cascade burst prediction and geographical distribution prediction, which will be explained below.

\subsection{Cascade Burst Prediction at Early Stages}
To predict the burstness of a cascade at early stages, we design a time-window structure by dividing the propogation time into different time windows and observe the diffusion process at the early time windows (no more than $Q$ time windows \footnote{Our model allows that observation window length and the maximum observation window number can be specified by the operator according to the service requirements. For example, in our experiment, a WM webpage cascade typically has a lifetime of more than one month. And hence we can set the observation window length as a day and the proposed algorithm requires no more than 5 observation windows to achieve high prediction accuracy.}) and conduct TW-LSTM method to do cascade burst prediction. In each time window $T^{k}$ ($k \leq Q$), we extract attributes from both macro and micro views. As we mainly describe the cascade diffusion analysis in one time window in the following discussion, to simplify the expression, we omit the superscript $k$ when it does not cause ambiguity.

\begin{figure*}[!t]
	\centering
	\includegraphics[width=0.9\linewidth]{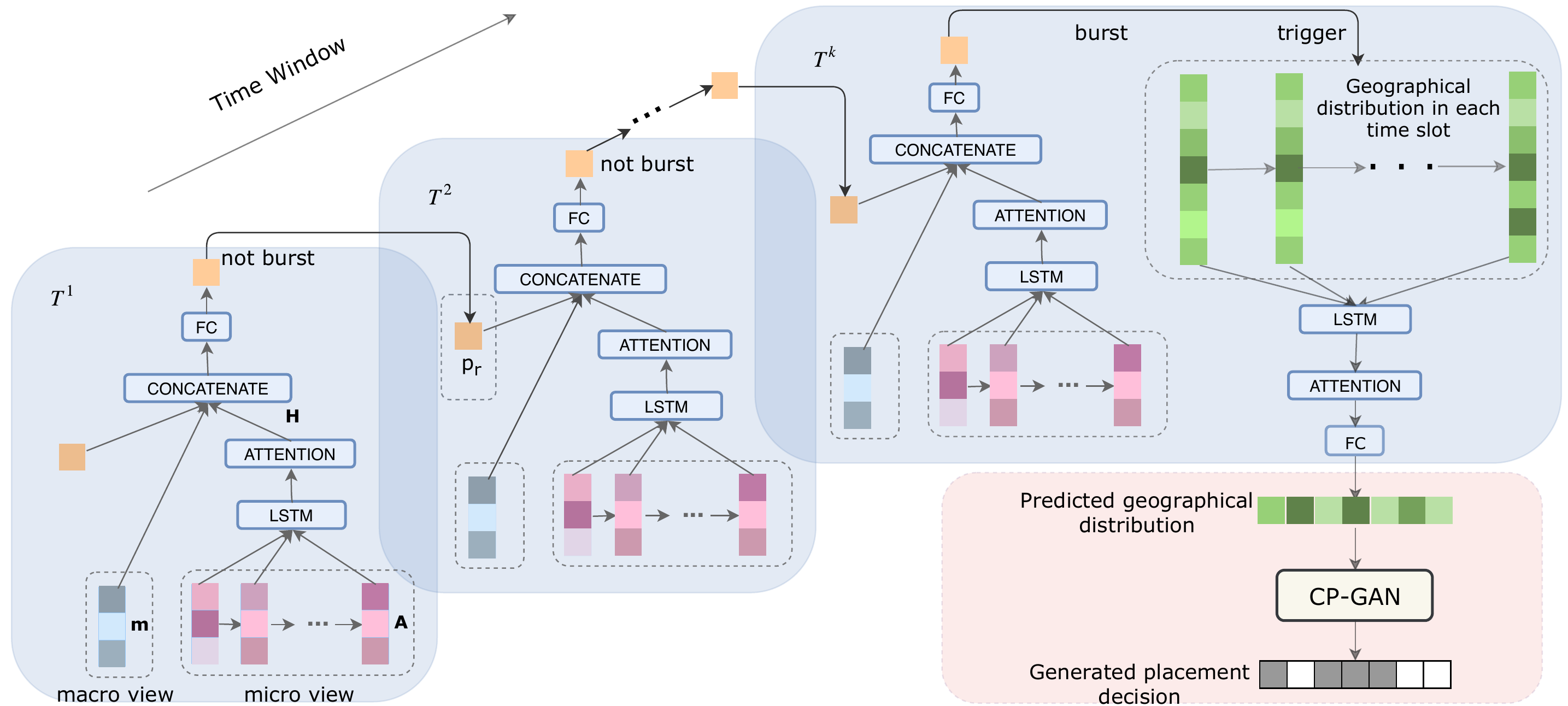}
	\caption{The DeepCP framework. TW-LSTM method which predict content popularity is shown in blue boxes. We observe the cascade at early several time windows and predict the burstness of the cascade. If the cascade is predicted to burst, we will evaluate the eventual geographical distribution of content popularity. The predicted results will further be used as inputs for autonomous content placement generation, which is illustrated in the pink box.}
	\label{Framework}
\end{figure*}

\begin{table*}[!t]
	\normalsize
	\caption{List of features extracted from macro view. }
	\label{featureList}       
	\centering
	\begin{tabular}{p{3cm} p{12cm}}
		\hline
			\multicolumn{2}{c}{\textbf{Infection perspective}}\\\hline
			$view\_count_{k}$ & the number of views in the $k$-th time window $T^{k}$\\\hline
			$reshare\_count_{k}$ & the number of reshares in the $k$-th time window $T^{k}$\\ \hline
			\multicolumn{2}{c}{\textbf{Topological perspective}}\\\hline
			$min\_Rlevel_{k}$ & the minimum level of reshares in the $k$-th time window $T^{k}$\\\hline
			$max\_Rlevel_{k}$ & the maximum level of reshares in the $k$-th time window $T^{k}$\\\hline
			$avg\_Rlevel_{k}$ & the average level of reshares in the $k$-th time window $T^{k}$\\\hline
			$var\_Rlevel_{k}$ & the variance for the levels of all reshares in the $k$-th time window $T^{k}$\\\hline
			$min\_Vlevel_{k}$ & the minimum level of views in the $k$-th time window $T^{k}$\\\hline
			$max\_Vlevel_{k}$ & the maximum level of views in the $k$-th time window $T^{k}$\\\hline
			$avg\_Vlevel_{k}$ & the average level of views in the $k$-th time window $T^{k}$\\\hline
			$var\_Vlevel_{k}$ & the variance for the levels of all views in the $k$-th time window $T^{k}$\\\hline
			\multicolumn{2}{c}{\textbf{Concent Type}}\\\hline
			$Ctype$ & the type of the spreading content, such as advertising, political pages\\\hline
			
	\end{tabular}
\end{table*}

\subsubsection{\textbf{Macro View}}
Based on the analysis in Section \ref{SectionAnalysis}, for a given time window $T$ at the early stage of the cascade diffusion process, the \emph{numbers of views and reshares} reflect the diffusion capacity of the cascade. The \emph{minimum, maximum and average levels of views and reshares} can imply the shape of the cascade tree, which may reflect the infection ability of the cascade. In addition, considered the observation in Section \ref{SectionAnalysis}, we also use \emph{content type} as a feature to do cascade burst prediction. Table \ref{featureList} describes the features which are extracted only from the cascade diffusion process without involving users' personal information and network correlations. We use ${m}$ to represent these features extracted from macro view in the learning model shown in Fig. \ref{Framework}.

\subsubsection{\textbf{Micro View}}
\label{microView}
To capture a more fine-grained cascade diffusion dynamics, we split a given observation time window $T$ into  $N$ time slots $T_{t}$ where $t = \{1, 2, \cdots, N\}$. For each time slot $T_{t}$, we extract some basic attributes (e.g., view number, reshare number, average level, etc.) and then the micro view features of the cascade diffusion process in time window $T$ can be represented as:
\begin{equation} \label{eqn3}
\begin{split}
{A} &= \{{a}_{1},{a}_{2}, \cdots,{a}_{N}\},\\
\end{split}
\end{equation}
where ${a}_{t}$ is the features in time slot $T_{t}$.

To represent the growth of cascades and learn from micro view features ${A}$ in a specific time window, we use Long Short-Term Memory (LSTM) network, which is known to be effective for modeling sequences \cite{Hochreiter1997Long}. LSTM is a special recurrent neural network (RNN) that remedies the vanishing gradient problem characteristic to RNNs in the long sequence training \cite{Gers2000Learning}. When applying it recursively to our cascade growth sequence ${A}$, LSTM can learn sequential correlations stably by maintaining a \textit{memory cell} ${c}_{t}$ in time slot $T_{t}$, which can accumulate the previous sequential information. In addition, LSTM introduces three gates: ${i}_{t}$, ${o}_{t}$, ${f}_{t}$, which decide the amount of new information to be added and the amount of history to be forgot and then the model can ``learn to forget'' in the temporal dimension. For a given time slot $T_{t}$, LSTM produces a hidden state ${h}_{t}$ based on the current input ${a}_{t}$ as well as the previous state ${h}_{t-1}$ and ${c}_{t-1}$. The architecture of LSTM can be formulated as follows:

\begin{equation} \label{eqn2}
\begin{split}
{i}_{t} &= \sigma({W}_{i}{a}_{t}+{U}_{i}{h}_{t-1}+{b}_{i}),\\
{f}_{t} &= \sigma({W}_{f}{a}_{t}+{U}_{f}{h}_{t-1}+{b}_{f}),\\
{o}_{t} &= \sigma({W}_{o}{a}_{t}+{U}_{o}{h}_{t-1}+{b}_{o}),\\
{\theta}_{t} &= tanh({W}_{a}{a}_{t}+{U}_{a}{h}_{t-1}+{b}_{a}),\\
{c}_{t} &= {f}_{t} \odot {c}_{t-1} + {i}_{t} \odot {\theta}_{t},\\
{h}_{t} &= {o}_{t} \odot tanh({c}_{t}).
\end{split}
\end{equation}
Here `$\odot$' denotes Hadamard product, $\sigma(\cdot)$ is sigmoid function and $tanh$ is hyperbolic tangent function. ${W}_{(\cdot)}$ and ${b}_{(\cdot)}$ denote weights and biases that are learned during model training.

Furthermore, we enforce the model to focus on important parts through an attention mechanism (see Fig. \ref{Framework}). Then the output of attention mechanism can be defined as:
\begin{equation} \label{eqn3}
\begin{split}
{H} &= \sum_{t=1}^{N}\alpha_{t}{h}_{t},\\
\end{split}
\end{equation}
where ${h}_{t}$ is the $t$-th hidden state of LSTM, N is the number of time slots,  and the attention coefficient $\alpha_{t}$ can be expressed as below:
\begin{equation} \label{eqn4}
\begin{split}
\alpha_{t} = softmax[tanh({W}_{\alpha}{h}_{t})+{b}_{\alpha}],\\
\end{split}
\end{equation}
where ${W}_{\alpha}$ is the weight matrix of attention layer and ${b}_{\alpha}$ is the bias.

\subsubsection{\textbf{Priori}}
If a cascade is predicted not to burst in a time window $T$, then we can continue to observe it in the next time windows to predict whether it will burst eventually. As we want to use learning results in previous time window for improving the prediction performance, we introduce a new parameter, priori $p_{r}$ in our model. As described in Fig. \ref{Framework}, the predicted burst probability during an observation time window can be used as a priori for the later time window. The priori for the first time window is set to be 0.5.

\subsubsection{\textbf{Prediction Component}}
Recall that our goal is to predict the burstness of a cascade, we join the features ${H}$ from micro view, ${m}$ from macro view and priori $p_{r}$ by concatenating them with the concatenation operator $\oplus$:
\begin{equation} \label{eqn_pc}
\begin{split}
{R} = {m}\oplus {H} \oplus p_{r}.\\
\end{split}
\end{equation}
Then we feed ${R}$ to the fully-connected neural network to get the burst prediction value $\hat{p}$ in a specific time window. The final prediction function can be defined as:
\begin{equation} \label{eqn_pc}
\begin{split}
\hat{p} = \sigma({W}_{fc}{R}+{b}_{fc}),\\
\end{split}
\end{equation}
where ${W}_{fc}$ and ${b}_{fc}$ are learnable parameters and $\sigma(\cdot)$ is sigmoid function.

Our proposed TW-LSTM method can learn the evolution process of the content cascade in a diffusion-aware manner by the new design of time-window structure and the sequential learning ability. During Q observation time windows, when the model predicts a cascade to burst, then we will launch the geographical distribution prediction component, which will be used for content placement decision generation later on.

\subsection{Geographical Distribution Prediction}
\begin{figure}[!t]
	\centering
	\subfigure[predicted geo-distribution]{
		\includegraphics[width=0.46\linewidth]{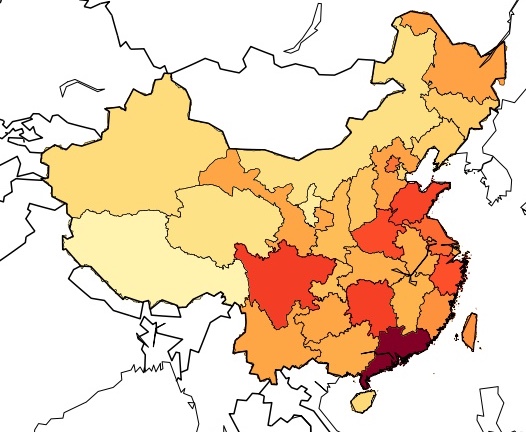}
	}
	\subfigure[real geo-distribution]{
		\includegraphics[width=0.46\linewidth]{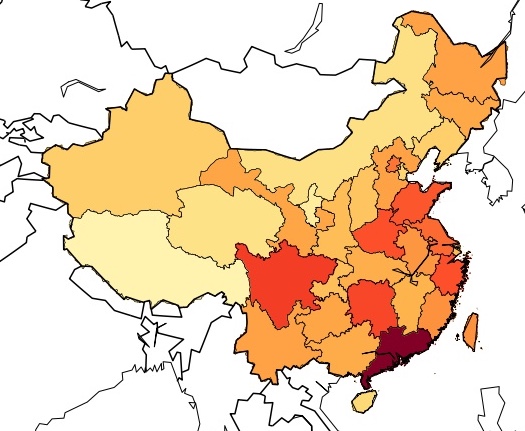}
	}
	\caption{An example of the comparison of our predicted geographical distribution with ground truth.}
	\label{compare_real_predict}
\end{figure}

To predict the eventual geographical distribution of content popularity, we observe the evolution of cascade diffusion in $M$ geographical regions only in the time window when we predict it to burst, since this time window contains the accumulated information of the infected users' geo-distribution up to now. For this time window, we also divide it into $N$ fine-grained time slots $T_{t}$ where $t = \{1, 2,..., N\}$ just as what we do in \ref{microView}. We observe the distribution of views and reshares in $M$ geographical regions in each time slot $T_{t}$, and then utilize LSTM model to explore the potential dependencies of these sequences containing infected users' geographical distributions. Moreover, attention mechanism is applied to allow the model to capture the most significant parts of the geographical distributions in $N$ time slots. Finally, we predict the eventual geographical distribution ${x}$ of the cascade by a stack of fully-connected layers. Although the eventual geo-distribution for different cascades presents significant diversity, our approach is able to achieve satisfactory results as it can learn from the evolution of cascade diffusion process. We give an example of the comparison between the predicted geographical distribution and the real geographical distribution for one WM page in Fig. \ref{compare_real_predict}, which shows that our predicted distribution is very close to the real one with only slight difference. More precisely, we quantify the accuracy of the predicted geographical distribution through the mean square error (MSE) as given below:
\begin{equation} \label{eqn6}
\begin{split}
&MSE =  \sum_{m=1}^{M}\frac{({x}_{m}-{\lambda}_{m})^{2}}{M},\\
\end{split}
\end{equation}
where ${x}_{m}$ is the predicted proportion of content popularity for region $m$ and ${\lambda}_{m}$ is ground truth. The average MSE for all WM pages is only 0.001, validating that our method can accurately predict the eventual geo-distribution of explosive cascades. With the predicted geographical distribution of content popularity, we can make content placement decision in order to reduce the content access latency and improve users' QoE.

\section{Autonomous Content Placement Decision Generation}
\label{SectionPlacement}
When a content cascade bursts, it is highly desirable for content provider to carry out content placement over several critical regions in order to reduce the content access latency by the massive geographically-diverse infected users and boost users' QoE. In general, the content placement problem is by nature a combinatorial optimization problem \cite{DBLP:conf/ifip5-5/DrwalJ12} and hence it is difficult to obtain the optimal solution in a real-time manner.

To address this issue, we devise CP-GAN model, a novel scheme for real-time content placement (CP) decision making by leveraging the power of generative adversarial network (GAN). Specifically, we first obtain the optimized decision samples by solving the content placement optimization in an offline manner. Then, using these optimal decision samples, we train the CP-GAN model for learning the optimal decision patterns and generating efficient online content placement decision in a real-time manner. Moreover, numerical results show that CP-GAN method is not sensitive to the parameter of maximum number of content replicas and outperforms existing approximation algorithms such as greedy scheme \cite{Salahuddin2015Social, Chen2012Intra, PapagianniLP13}.

\subsection{Offline Optimal Decision Sample Acquisition}
In this part, we first obtain the offline optimal content placement decision samples by formulating and solving the content placement problem in an offline manner. Suppose that users are from $\mathcal{M} = \{1, 2, \cdots, M\}$ regions and there are $S_{i}$ users who view or reshare the WM page in region $i \in \mathcal{M}$. To shorten the communication distance between content and users, we can place content replicas over the critical regions. However, due to operational cost constraint (e.g., content delivery network resource usage budget limit), it is typically impossible to place the content over all the regions. Thus, we introduce $C$ to denote the maximum number of replicas for a content. Here we take a binary indicator $I_{j}$ to denote the content placement decision variable in region $j$, let $I_{j} = 1$ if the content is placed in region $j$ and $I_{j} = 0$ else. Therefore, our problem of minimizing the content access latency of all infected users involved in a cascade can be formulated as the following optimization:
\begin{equation} \label{eqn5}
\begin{split}
&\min_{ v_{ij}, I_j } \quad \sum_{i=1}^{M}\sum_{j=1}^{M}v_{ij}l_{ij}
\end{split}
\end{equation}
\begin{align}
s.t. \quad &\sum_{i=1}^{M}v_{ij} \le I_{j}U_{j}, \\
&\sum_{j=1}^{M}v_{ij} = S_{i}, \\
&\sum_{j=1}^{M}I_{j} \le C, \\
&I_{j} \in \{0,1\},
\end{align}
where $v_{ij}$ indicates the number of infected users in region $i$ who get content from replication content servers in region $j$. $l_{ij}$ is the access delay from region $i$ to region $j$. (9) states that the number of users served by the content servers in region $j$ is no more than its service capacity $U_{j}$ if there is a content replication in region $j$. Otherwise, users in region $i$ will not get content from region $j$. (10) indicates that the number of served users in region $i$ equals to the infected users in region $i$. (11) represents the constraint of the maximum number of allowable content replicas. For simplicity, we only consider the capacity constraint. Our method can also apply in other cases with more system constraints by only feeding those constraint-feasible samples and decisions generated offline to CP-GAN.

The content placement optimization problem in (8) is a mixed integer programming (MIP) problem. In practice, it is difficult to find the optimal solution in a fast manner. Nevertheless, in our solution, we would like to generate optimized decision samples and this can be conducted in the offline manner. And hence we solve problem in (8) using the MIP solver---IBM CPLEX Optimizer \cite{kellerer2004multidimensional}.

\subsection{Autonomous Content Placement Decision}
Based on the obtained optimal decision samples in the previous offline stage, we can utilize the generative adversarial network (GAN) to learn from the offline optimal content placement samples and approximate the offline optimal decisions. By using the trained GAN model, we can then generate efficient online content placement decision in a real-time manner.

Generative adversarial networks (GANs), first proposed by Goodfellow et al. \cite{Goodfellow2014Generative}, set up an adversarial game between a discriminator network $D$ and a generator network $G$. The goal of $G$ is to generate samples that are not distinguishable by $D$, while $D$ estimates the probability that a sample comes from the training data rather than generated from $G$. Through alternating adversarial training, GAN ia able to achieve excellent sample generation ability and has been successfully applied to many applications, such as image generation task. And later, GAN has been extended for semi-supervised learning and validated to be effective \cite{Salimans2016Improved}.

In our problem, for the predicted geo-distribution $\mathbf{x}$ of each explosive WM page, we want to generate its content placement strategy $\mathbf{y}$ which is a binary vector and selects up to $C$ regions to place content according to the amount of infected users in all $M$ regions. We first use offline placement decision method to get the optimal placement decisions for a few WM pages. Therefore, we have some limited labeled data available for the training of GAN. We also add samples generated from G using random vector $z$ to our dataset. Correspondingly, the dimension of our discriminator output $\mathbf{y}$ increases from $M$ to $M+1$ and it means that if the last dimension $\mathbf{y}_{M+1}$ is equal to one, then the input is generated from G. The offline training procedure is illustrated in Fig. \ref{offline_placement}. We use $p_{model}(\mathbf{y}_{1,2,...,M}|\mathbf{x})$ to represent the probability that the placement strategy is $\mathbf{y}_{1,2,...,M}$ and $p_{model}(\mathbf{y}_{M+1}|\mathbf{x})$ to supply the probability that  $\mathbf{x}$ is fake. And it also suggests that the content placement decision is a feasible solution if $\mathbf{x}$ is true. Our goal for supervised learning is to minimize the total cross-entropy loss between the optimal content placement and our generated placement decision. For unsupervised learning, we train D to maximize the probability of assigning the correct label to both real samples and generated samples in our data set. Assuming that half of our data set is real data and half is generated from $G$, our loss function for training discriminator $D$ contains two parts: supervised loss function $L_{s}$ and unsupervised loss function $L_{u}$, and can be formulated as follows:

\begin{figure}[!t]
	\centering
	\includegraphics[width=0.95\linewidth]{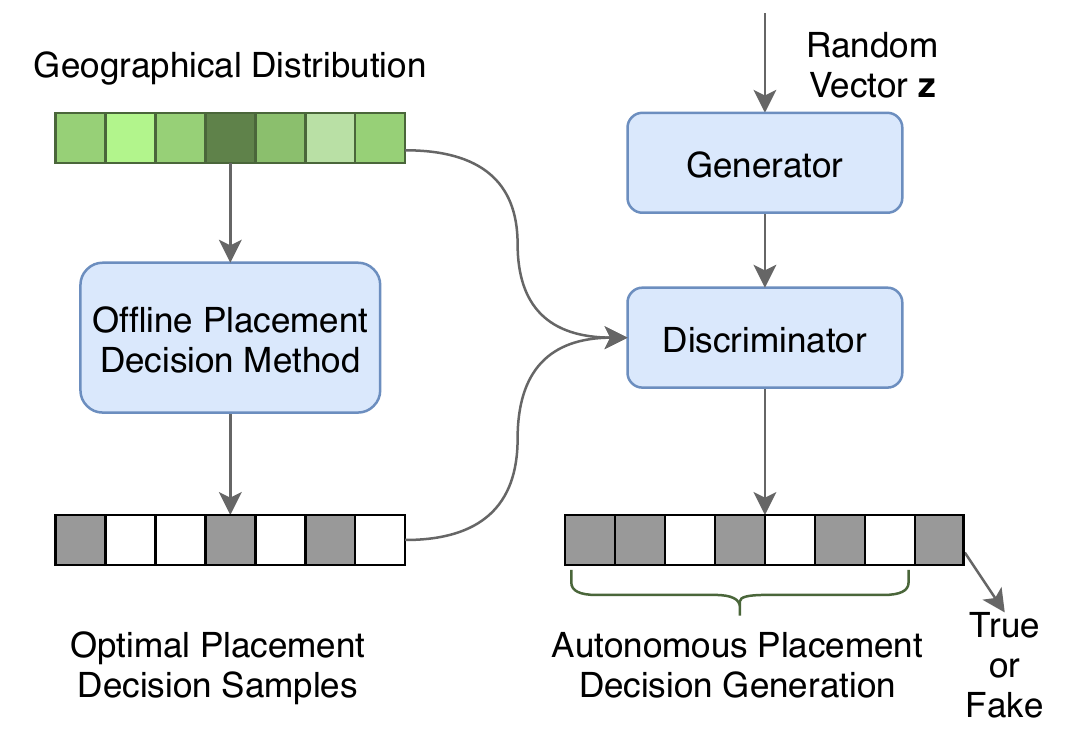}
	\caption{The offline training procedure using CP-GAN for autonomous content placement decision generation.}
	\label{offline_placement}
\end{figure}

\begin{equation} \label{eqn6}
\begin{split}
&L =  L_{s} + L_{u},\\
\end{split}
\vspace{-25pt}
\end{equation}
where
\begin{align}
L_{s} &=  - \mathbb{E}_{\mathbf{x,y}\sim p_{data}(\mathbf{x,y})}\log p_{model}(\mathbf{y}|\mathbf{x},\mathbf{y}_{M+1}=0),
\end{align}
\begin{align}
\begin{gathered}
L_{u} = -\{\mathbb{E}_{\mathbf{x}\sim p_{data}(\mathbf{x})}\log[1-p_{model}(\mathbf{y}_{M+1}=1|\mathbf{x})] \\
+ \mathbb{E}_{\mathbf{x}\sim G}\log[p_{model}(\mathbf{y}_{M+1}=1|\mathbf{x})]\}.
\end{gathered}
\end{align}

As deep neural network (DNN) has been identified as a universal function approximator that can approximate any function mapping, possibly with multiple inputs and outputs \cite{HornikSW89}, we use DNN as the networks for both the generator and the discriminator of GAN for optimal content placement samples learning and content placement decision making. The networks for generator and discriminator of GAN are both implemented as a 3 layer deep neural network (DNN). Specifically, the number of neurons in generator is 32, 64 and 128 for each layer, while the number of neurons in each layer of discriminator is 64.

\begin{figure}[!t]
	\centering
	\includegraphics[width=0.95\linewidth]{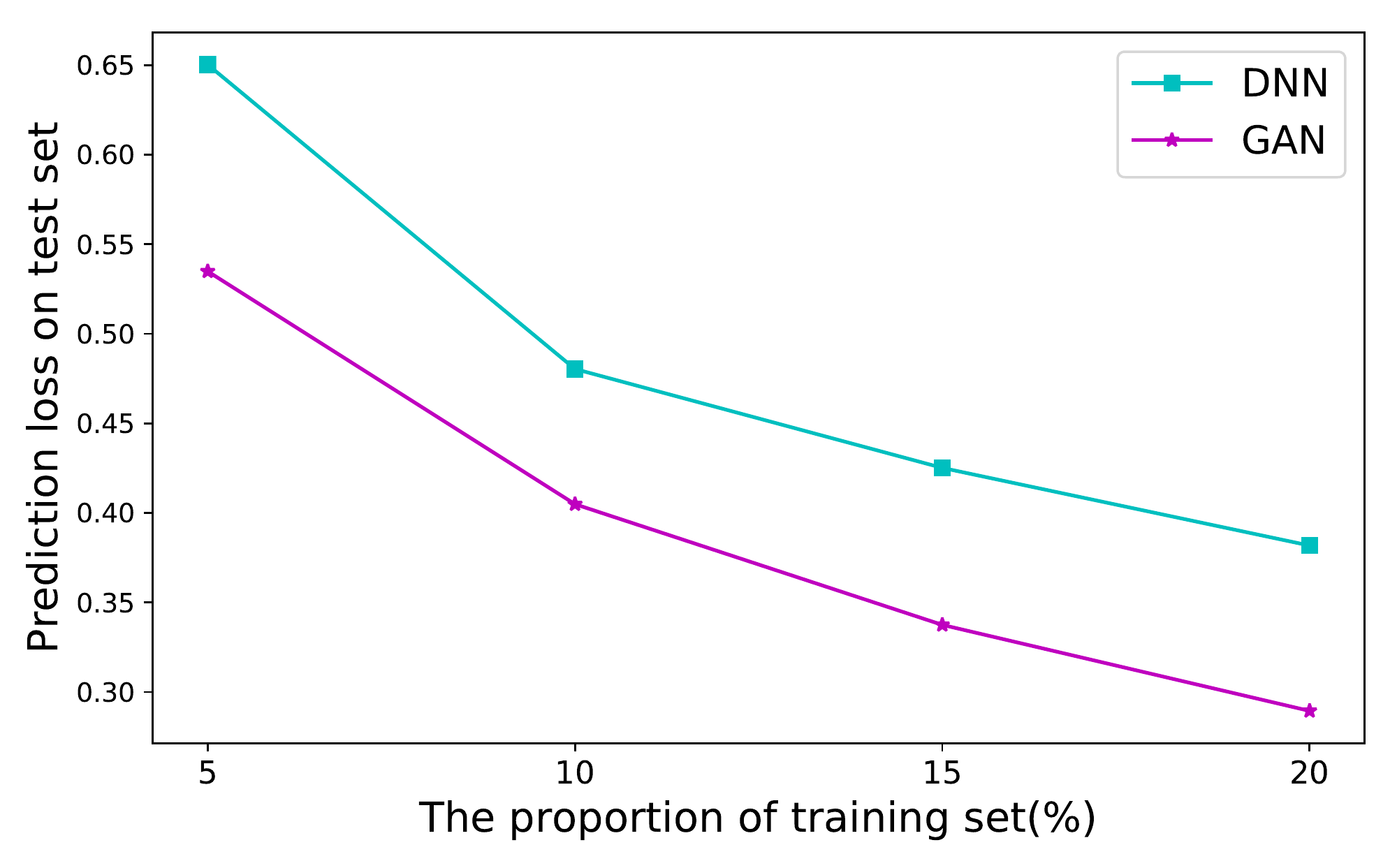}
	\caption{The prediction loss on test set with different proportions of training set. GAN can achieve lower loss comparing with DNN.}
	\label{gandnn}
\end{figure}

GAN is an ideal model to well approximate the optimal content placement strategy for the following two reasons: First, GAN can learn from the offline optimal content placement samples and then make placement decisions for new contents in a fast manner, without solving the optimization problem in (8). GAN takes the content popularity distribution as input and outputs the probability of placing content for each region. We select the top $C$ regions to replicate content. Second, the goal of G is to generate samples for D to alleviate the problem of limited labeled samples, while D is used to estimate the content placement strategy, that is, to predict the content caching probability for each region. Thus, to evaluate the content placement decision making performance, we compare GAN with the pure DNN, a widely adopted prediction neural network that has no ability to generate new samples for model training and is directly trained using limited offline optimal decision samples. As depicted in Fig. \ref{gandnn}, with limited offline optimal content placement samples, GAN is able to achieve lower prediction loss when making content placement decision comparing with DNN which has the same structure with the discriminator of GAN. It suggests that unlike DNN, GAN does not require a very large sample size, which justifies our choice of GAN.

Based on the trained CP-GAN model, as introduced in Fig. \ref{Framework}, when a WM page is predicted to burst, the eventual geographical distribution prediction will be triggered, and then the predicted geo-distribution of content popularity will be injected into CP-GAN model which can generate the real-time content placement decision over the regions. Our experiments show that the proposed GAN-based content placement decision generation mechanism can well approximate the optimal solution and achieve real-time decision generation.

\section{Performance Evaluation}
\label{SectionExper}
In this section, we demonstrate the effectiveness of our proposed DeepCP framework. Specifically, we evaluate the performance on diffusion-aware cascade prediction and autonomous content placement decision generation, separately. The dataset used is described in Section \ref{SubsectionDataset}.
\subsection{Cascade Prediction Evaluation}
\subsubsection{\textbf{Hyperparameter Setting}}
For cascade prediction, our TW-LSTM method is implemented as a three-layer network composed of one LSTM layer with 64 hidden neurons and two fully connected layers with 32, 16 neurons, respectively. We set the time window $T=24$ (e.g., 24 hours) and the sequence length $N=24$ for LSTM. To train the proposed TW-LSTM model, we use TensorFlow platform and backpropagation algorithm with Adam optimizer, which commonly yields faster convergence compared to traditional stochastic gradient descent (SGD) \cite{DBLP:journals/corr/KingmaB14}.

\subsubsection{\textbf{Comparing Methods for Cascade Prediction}}
We compare our TW-LSTM cascade prediction method with both statistic method and machine learning classifiers widely used in \cite{EnhancingWang} and \cite{HornikSW89}.  For machine-learning based methods, we use the same features and tune the parameters for all methods. Besides, for each time window, we conduct cascade burst predicition and feed the predicted burst probability to the next time window as a priori just as TW-LSTM does.

\begin{itemize}
	\item \textbf{Holt-Winters Exponential Smoothing (HW-ExpS)} \cite{natrella2010nist} is a quantitative forecasting method that uses mathematical recursive functions to predict the trend behavior. We use the propagation series in the first few days to predict the eventual cascade size in the end of the month.
	\item \textbf{Support Vector Machine (SVM)} is a popular classifier that is demonstrated to be effective on a huge category of classification problems.
	\item \textbf{Logistic Regression (LR)} is a statistical model that is usually taken to apply to a binary dependent variable, which represents whether the cascade will burst or not in our study.
	\item \textbf{Neural Networks (NN)} have been proven effective for time series prediction \cite{azoff1994neural} and are applied to predict the number of views for a video content in \cite{EnhancingWang}. We use the same three-layer feed-forward neural network as in \cite{EnhancingWang} and the number of neurons for each layer is 64, 32 and 16.

	\item \textbf{Random Forest (RF)}  is an ensemble learning method by building a set of decision trees with random subsets of attributes and bagging them for classification results.
	\item \textbf{Gradient Boosting Decision Tree (GBDT)} produces a prediction model in the form of an ensemble of weak prediction models and generalizes them by allowing optimization of an arbitrary differentiable loss function.
\end{itemize}

\begin{table}[!t]
	\caption{Performance with different number of time windows. }
	\label{results}       
	\centering
	\begin{tabular}{|c|c|c|c|c|c|}
		\hline
		Method(\%) & $T^{1}$ & $T^{2}$ & $T^{3}$ & $T^{4}$ & $T^{5}$ \\
		\hline HW-ExpS & $59.26$ & $63.54$ & $63.71$ & $69.68$ & $63.37$ \\
		\hline SVM & $57.76$ & $51.26$ & $60.40$ & $69.61$ & $72.56$\\
		\hline LR & $53.57$ & $52.10$ & $61.45$ & $69.47$ & $73.58$ \\
		\hline NN & $75.25$ & $69.60$ & $74.65$ & $78.81$ & $79.62$\\
		\hline RF & $79.43$ & $81.84$ & $80.33$ & $82.81$ & $84.35$\\
		\hline GBDT & ${79.80}$ & $79.05$ & $79.21$ & $81.90$ & $82.52$\\
		\hline\hline TW-LSTM-t-p & $74.51$ & $70.27$ & $78.35$ & $82.69$ & $85.07$ \\	
		\hline TW-LSTM-t & $79.41$ & $80.39$ & $81.19$ & $82.86$ & $86.49$ \\	
		\hline TW-LSTM & $79.61$ & ${83.41}$ & ${85.75}$ & ${87.09}$ & ${89.32}$\\
		\hline
	\end{tabular}
\end{table}

\subsubsection{\textbf{Cascade Prediction Performance}}
In order to evaluate the effectiveness of our proposed TW-LSTM model, we predict whether a cascade can burst by observing different numbers of time windows. We summarize these results in TABLE \ref{results}. When we predict whether a cascade will burst with the observation in only one time window, GBDT can achieve an accuracy of $79.80\%$ and our proposed TW-LSTM method can also perform well with the accuracy of $79.61\%$. When we observe more time windows, the TW-LSTM model can well outperform other models. Besides, we can find that our proposed method can bring the accumulation of prediction accuracy with the increasing number of observation time windows, while other models don't. For example, the prediction accuracies of SVM, LR and NN have declined to varying degrees in the second time window $T^{2}$.

To better understand the effect of the priori and content type, we also provide the experimental results. If we consider the priori in previous time window (TW-LSTM-t), the prediction accuracy can increase compared with TW-LSTM-t-p which does not consider the influence of previous prediction result. Moreover, we can see that the accuracy of TW-LSTM-t method grows with the increase of time windows, which shows that the priori is of great importance for cascade burst prediction. As for the content type, it is obvious that the prediction accuracy improves if we consider the effect of content type. For instance, TW-LSTM method can achieve an accuracy of $85.75\%$ in $T^{3}$ while if we ignore content type (TW-LSTM-t), the accuracy is $81.19\%$.

\subsection{Content Placement Decision Evaluation}
With the predicted geo-distribution of infected users, our proposed CP-GAN mechanism can make excellent placement decision to enhance users' QoE.

\subsubsection{\textbf{Baselines for Content Placement}}

\begin{itemize}
	\item \textbf{No replicating.} If the content is not replicated on servers in other popular areas, all infected users will get it from the server where FIBODATA is located.
	\item \textbf{Heuristic Placement.} In \cite{EnhancingWang}, authors propose a heuristic algorithm to replicate videos based on prediction. The key idea is to serve users locally if there is video content in current region, or to send users' requests to other regions, as assuming that users' region preference is uniformly distributed.
	\item \textbf{Distance-aware Placement.} If we don't know how popular the content is in each region, we can assume that the content is equally popular in each region and then select critical regions in which the content should be placed according to the distances between regions.
	\item \textbf{Statistics-based Placement.} According to the predicted eventual geographical distributions of all historical WM pages, we can obtain the statistical geographical distribution of all the pages and we use the statistical distribution information to solve the optimal placement problem.	
	\item \textbf{Greedy Placement.} With the predicted geo-distribution of infected users for a single content, we can select the first $C$ regions sequentially in a greedy manner that can significantly reduce the total access delay. The greedy scheme is widely-used in literature such as \cite{Salahuddin2015Social, Chen2012Intra, PapagianniLP13}.
\end{itemize}

\begin{figure}[!t]
	\centering
	\includegraphics[width=0.95\linewidth]{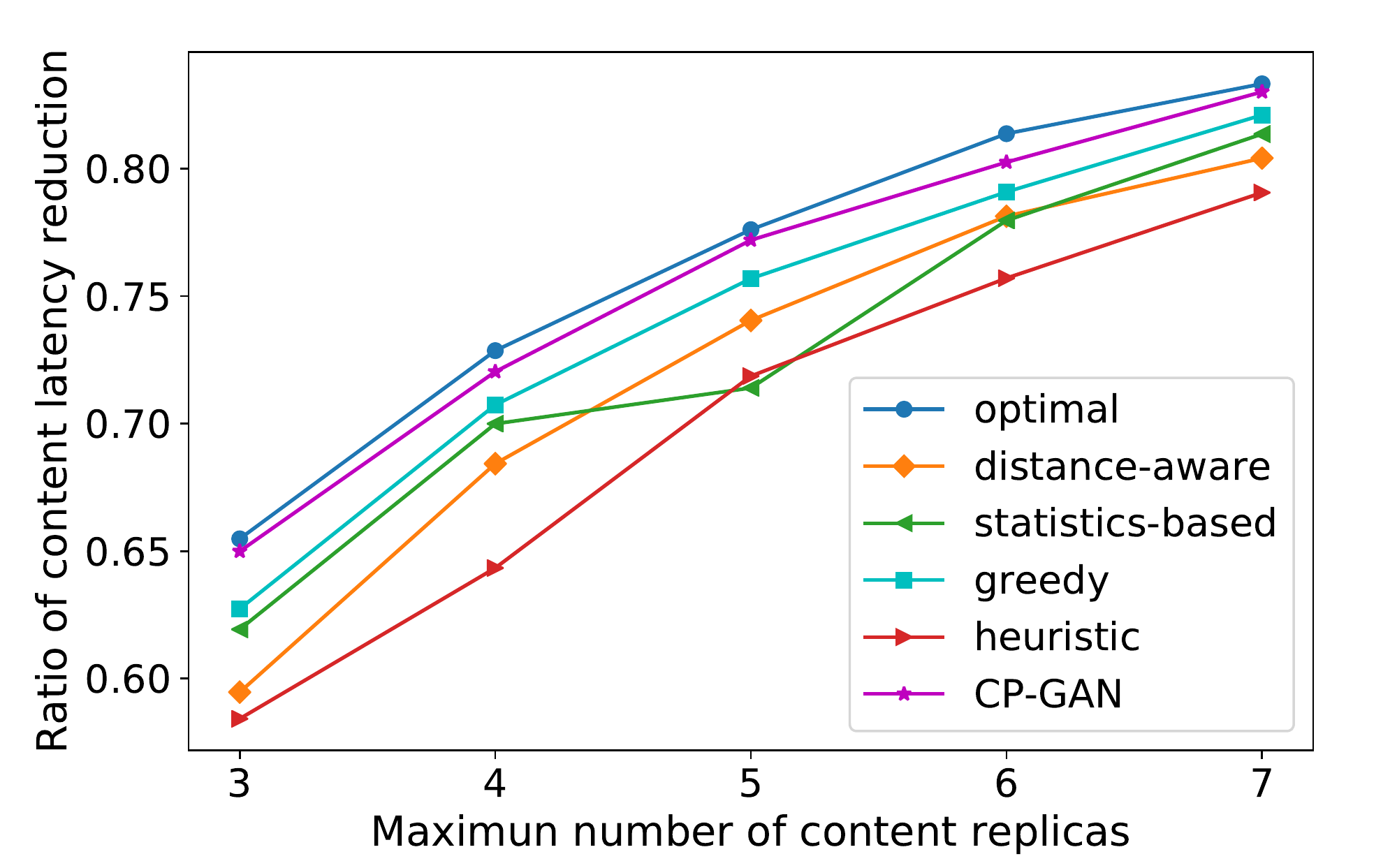}
	\caption{The ratio of content access latency reduction for different placement methods comparing with no replicating under different number of content replicas. Our proposed CP-GAN method can effectively reduce content access latency.}
	\label{total_cost}
\end{figure}

\subsubsection{\textbf{Content Placement Performance}}
In our experiment, we set the total regions $M = 34$ which is the number of provinces in China. For the maximum number of replicas $C$,  we choose different values to verify the performance of CP-GAN method. We take the content access latency caused by the placement of no replicating as benchmark, and use the ratio of content latency reduction brought by different algorithms compared to no replicating as the evaluation indicator. Experimental results are illustrated in Fig. \ref{total_cost}. Compared with other methods, CP-GAN method can effectively reduce the total access latency and is very close to the optimal placement, whatever the maximum number of content replicas. It is mainly because that CP-GAN is able to make wise content placement from a globle perspective after training with limit optimal decision samples. While the comparing methods (such as heuristic placement) make decisions only by the current interests and fails to consider the whole.

\begin{figure}[!t]
	\centering
	\includegraphics[width=0.95\linewidth]{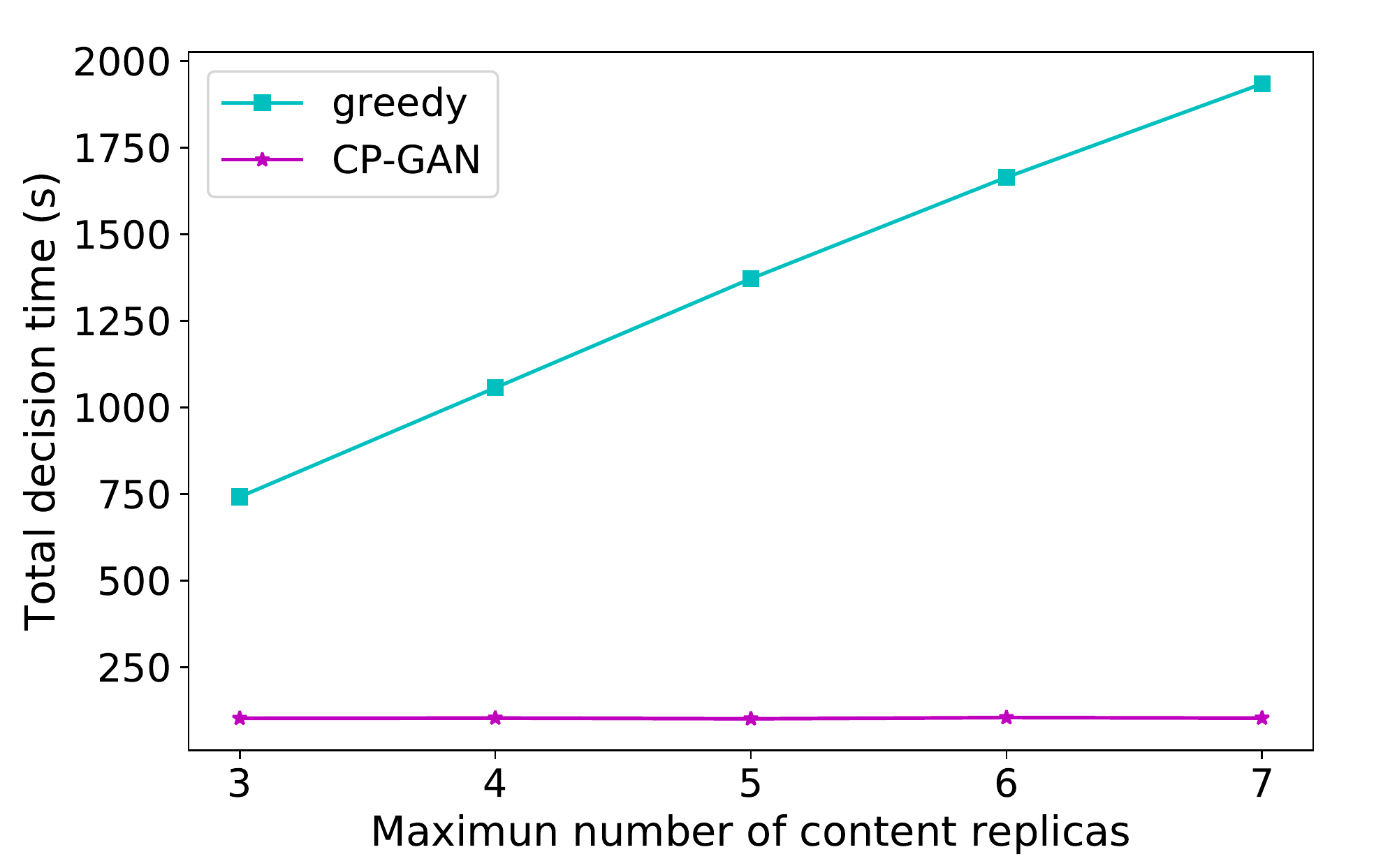}
	\caption{Total decision time needed for content placement decision generation for all pages under different number of content replicas. Our proposed method is much faster than greedy algorithm.}
	\label{total_time}
\end{figure}

Although greedy algorithm can generate good content placement decision, the decision time it takes is much longer than that of CP-GAN as shown in Fig. \ref{total_time}. Furthermore, the running time of greedy algorithm grows almost linearly as the maximum number of content replicas increases, since greedy algorithm needs to search over the solution optimization space for content placement decision making. While CP-GAN method is not sensitive to the parameter of maximum number of content replicas, as CP-GAN outputs the replication probability for each region with the learned policy at the online inference stage, no matter what the maximum number of content replicas is. And CP-GAN selects the top-$C$ regions with the highest probability to conduct content placement according to the maximum number $C$. This shows that our algorithm can make efficient decisions in a very short time, which is significant in the social network scenario where cascades are fast-spreading.

\begin{figure}[!t]
	\centering
	\includegraphics[width=0.95\linewidth]{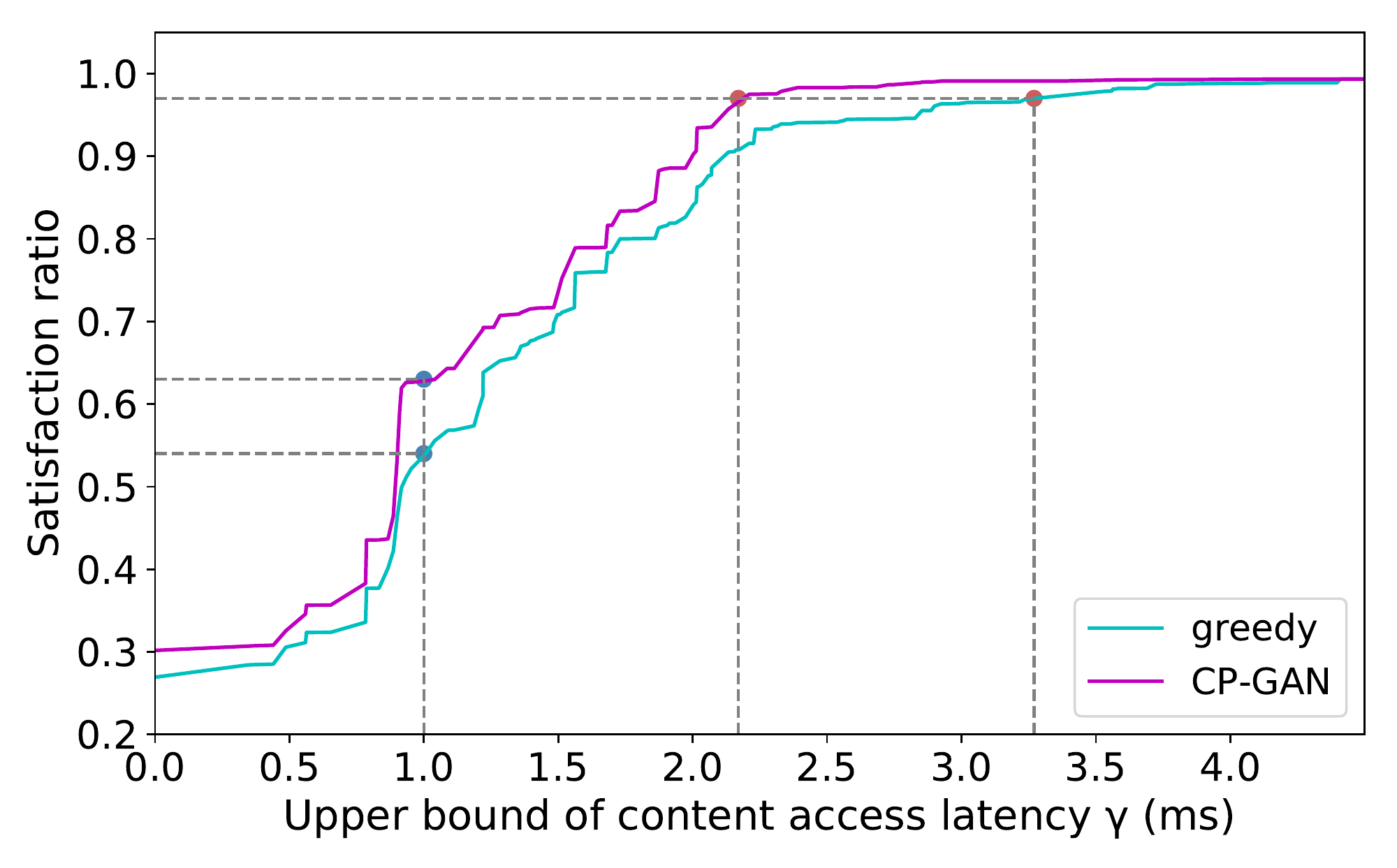}
	\caption{Satisfaction ratio under different upper bound of content access latency $\gamma$ when the maximum number of content replicas is 5.}
	\label{cdf}
\end{figure}

In order to show how CP-GAN method can effectively enhance user's QoE, satisfaction ratio is defined to represent the probability of users with satisfactory QoE and can be  formulated as follows:
\begin{equation} \label{eqn16}
\begin{split}
&satisfaction \; ratio =  Pr\{latency \leq \gamma\},\\
\end{split}
\end{equation}
where $\gamma$ is the upper bound of the latency that users can tolerate. Fig. \ref{cdf} illustrates the satisfaction radio under different values of $\gamma$ when the maximum number of content replicas is 5. When $\gamma$ is 1ms, more than $60\%$ of users can be served with satisfactory QoE under CP-GAN strategy, while less than $55\%$ if we use greedy method to conduct content placement. Moreover, almost all users can get access to the preferred content within less than 2.2ms in our CP-GAN scheme. Nevertheless, the upper bound  $\gamma$ grows to 3.3ms in greedy scheme. It shows that CP-GAN can make efficient and satisfactory content placement decisions and enhance users' QoE compared with the greedy algorithm.

\begin{figure}[!t]
	\centering
	\includegraphics[width=0.95\linewidth]{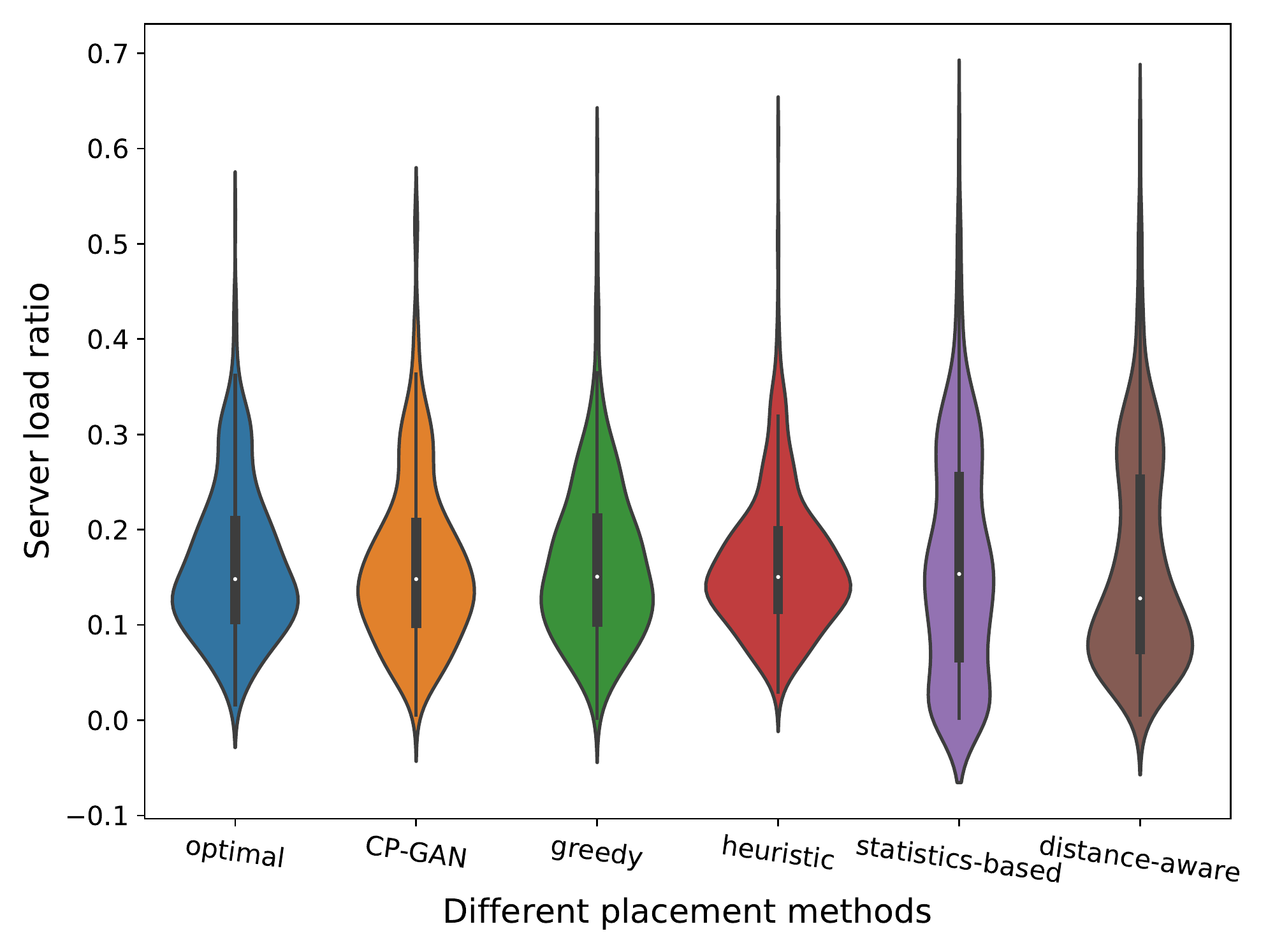}
	\caption{The distribution of server load ratio of 100 WM pages under different content placement approaches when the maximum number of content replicas is 5.}
	\label{server_load}
\end{figure}

Furthermore, we evaluate the server load ratio of different content placement approaches as depicted in Fig. \ref{server_load}. The distribution of server load ratio (the percentage of users served by the server ) for 100 WM pages under the content placement decision making strategy of CP-GAN is very similar with that of the optimal placement strategy, which indicates that CP-GAN can make near-optimal placement decision strategy comparing with other approaches.

\begin{figure*}[!t]
	\centering
	\subfigure[optimal placement]{
		\includegraphics[width=0.28\linewidth]{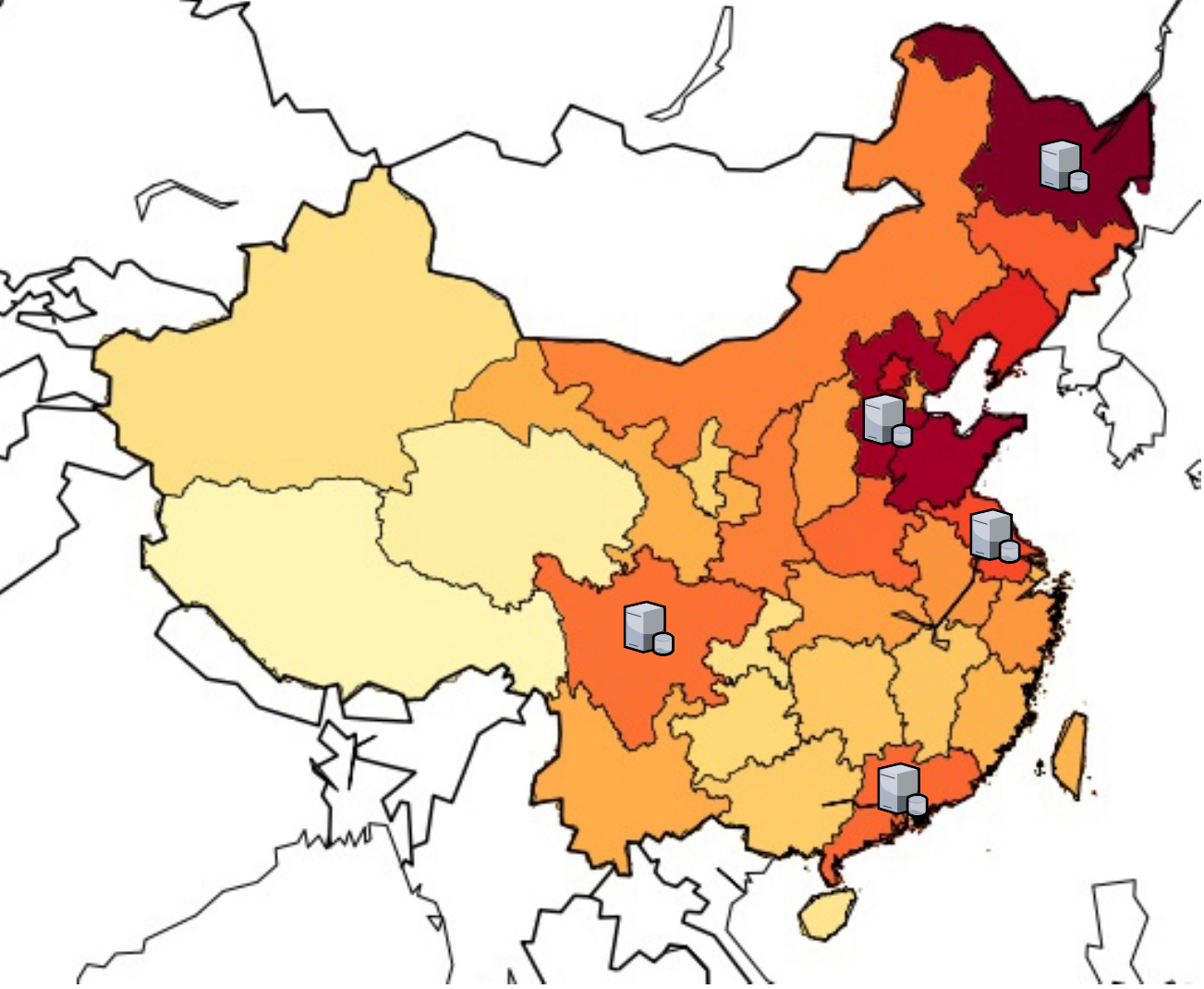}
	}
	\subfigure[CP-GAN placement]{
		\includegraphics[width=0.28\linewidth]{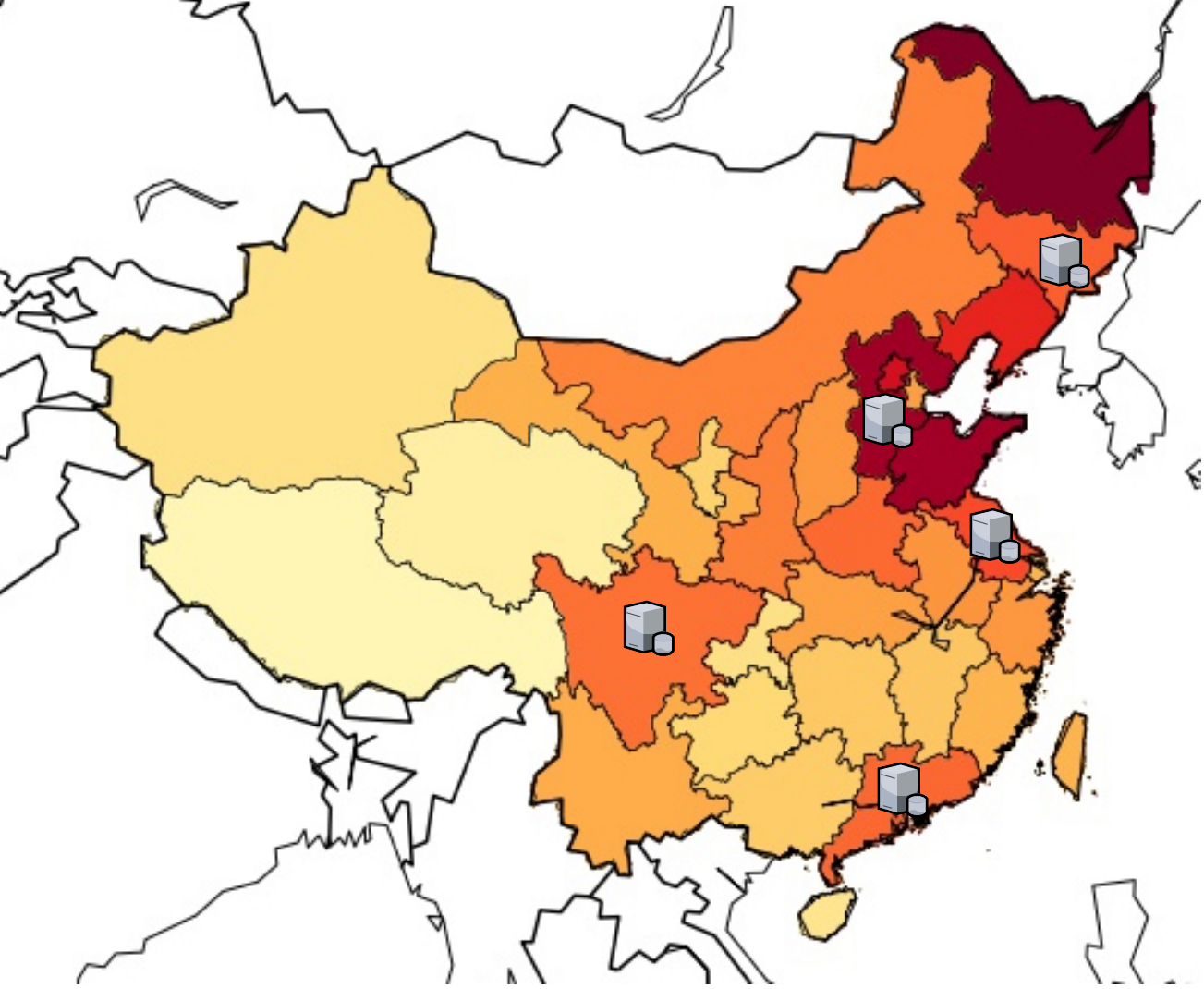}
	} 	
	\subfigure[greedy placement]{
		\includegraphics[width=0.28\linewidth]{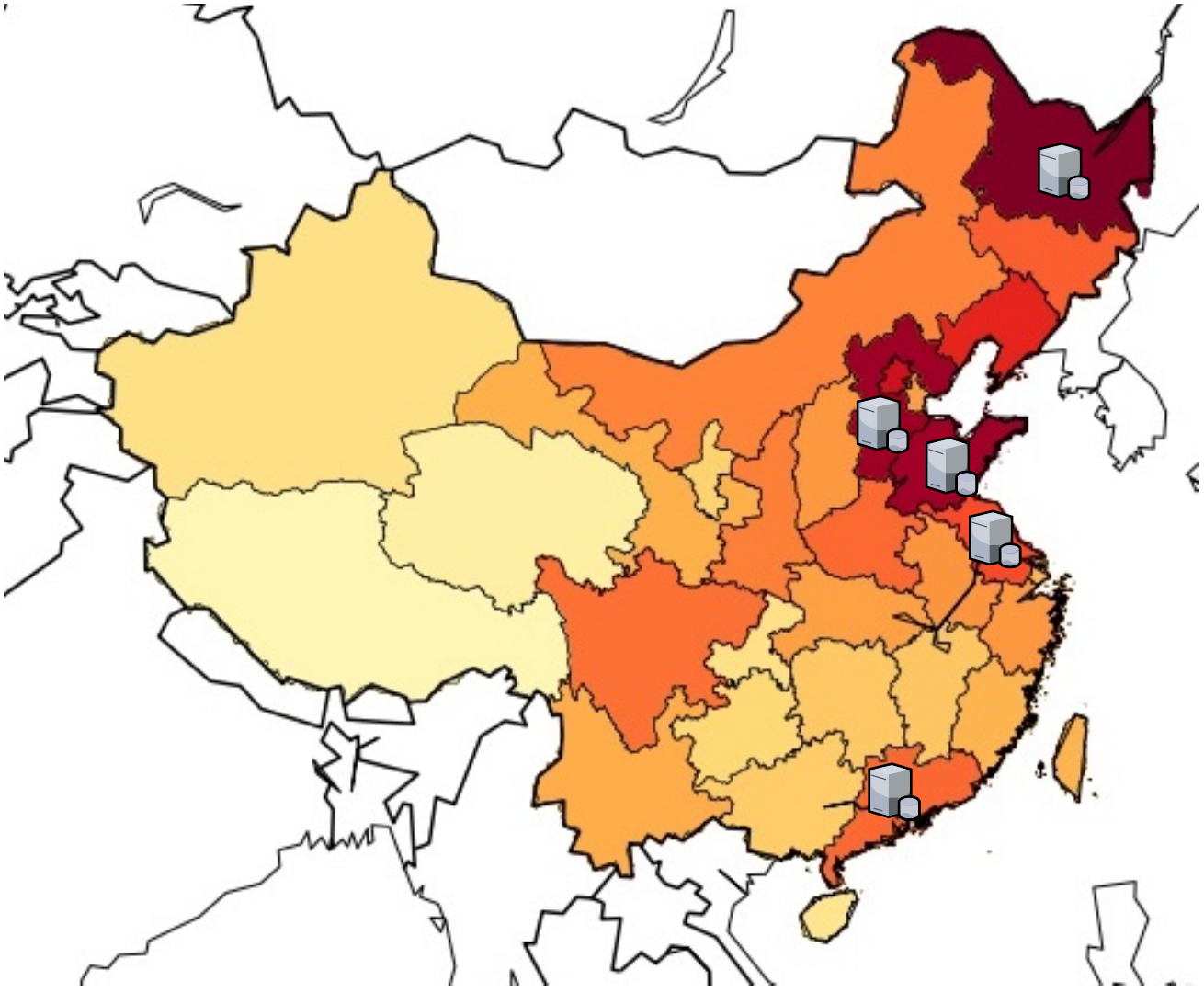}
	}
	\caption{Content placement decision of different methods for a given WM page if the maximum number of content replicas is 5. (a) represents the optimal placement which can minimize the total access latency of all infected users in different regions. (b) shows the placement decision of our proposed CP-GAN method and (c) is the placement of greedy algorithm.}
	\label{placement decision}
\end{figure*}

To better demonstrate the effectiveness of our proposed placement strategy, we provide a visualization of the comparison among optimal, greedy and CP-GAN methods for a given WM page replicating in 5 regions in Fig. \ref{placement decision}. Our content placement decision is quite similar to the optimal solution while the greedy placement decision fails to consider the nationwide placement of content.

\section{Conclusion}
\label{SectionConc}
Aiming at designing an efficient popularity-aware content placement mechanism in closed social networks, in this paper we propose a privacy-friendly framework of DeepCP which unifies diffusion-aware cascade prediction and autonomous proactive content placement. We utilize the toolset of deep learning to devise TW-LSTM and CP-GAN mechanisms accordingly. We conduct experiments using a real-world dataset, and extensive experiments show that our model can achieve a superior cascade prediction performance. And the generated content placement decision can also significantly reduce the access latency of users, which can greatly enhance users' QoE. We hope this work can help to stimulate a new line of thinking on leveraging the data-driven deep learning toolset for designing future-generation intelligent OSN content service systems.

\ifCLASSOPTIONcaptionsoff
  \newpage
\fi



%
\bibliographystyle{unsrt}
\bibliography{reference}

\end{document}